\documentclass[12pt]{article}

\usepackage{cite}
\usepackage{epsfig}
\usepackage{amsmath}
\usepackage{array}
\usepackage{axodraw}
\usepackage{mathbbol}

\def\mathswitch#1{\relax\ifmmode#1\else$#1$\fi}
\def\mathswitchr#1{\relax\ifmmode{\mathrm{#1}}\else$\mathrm{#1}$\fi}
\newcommand{\PW}{\mathswitchr W}

\newcommand{\scrs}{{}}
\newcommand{\sw}{\mathswitch {s_{\scrs\PW}}}
\newcommand{\cw}{\mathswitch {c_{\scrs\PW}}}

\newcommand{\gev}{\,\, \mathrm{GeV}}

\newcommand{\SLASH}[2]{\makebox[#2ex][l]{$#1$}/}

\newcommand{\Eslash}{\SLASH{E}{.3}\,}

\newcommand{\anc}{\rule{0mm}{0mm}}

\newcommand{\gesim}{\,\raisebox{-.3ex}{$_{\textstyle
>}\atop^{\textstyle\sim}$}\,}

\newcommand{\eps}{\epsilon}

\newcommand{\mycaption}[1]{\caption{\sl #1}}

\begin{document}
\thispagestyle{empty}

\def\thefootnote{\fnsymbol{footnote}}

\begin{flushright}
ZU-TH 08/08
\end{flushright}

\vspace{1cm}

\begin{center}

{\Large\sc {\bf Consequences of T-parity breaking\\ in the Littlest Higgs
model}}
\\[3.5em]
{\large\sc 
A.~Freitas$^1$, P.~Schwaller$^2$,
D.~Wyler$^2$
}

\vspace*{1cm}

{\sl $^1$
Department of Physics \& Astronomy, University of Pittsburgh,\\
3941 O'Hara St, Pittsburgh, PA 15260, USA
}
\\[1em]
{\sl $^2$
Institut f\"ur Theoretische Physik,
        Universit\"at Z\"urich, \\ Winterthurerstrasse 190, CH-8057
        Z\"urich, Switzerland
}

\end{center}

\vspace*{2.5cm}

\begin{abstract}

In this paper we consider the effects of the T-parity violating 
anomalous Wess-Zumino-Witten-Term in the Littlest Higgs model. Apart from tree
level processes, the loop induced decays of the heavy mirror particles into
light standard model fermions lead to a new and rich phenomenology in particular
at breaking scales $f$ below 1 TeV. Various processes are calculated and their
signatures at present and future colliders are discussed. As a byproduct we find an alternative production mechanism for the Higgs boson.

\end{abstract}

\def\thefootnote{\arabic{footnote}}
\setcounter{page}{0}
\setcounter{footnote}{0}

\newpage


\section{Introduction}

Little Higgs models provide a mechanism to explain a hierarchy between the
electroweak scale and a larger, fundamental scale where symmetry breaking occurs
through strong dynamics.
In this scheme, the Higgs scalar doublet is a composite particle of the strong
dynamics, a pseudo-Goldstone boson stemming from the spontaneously
broken symmetry at a scale $f$. The Goldstone mechanism protects the Higgs
boson from acquiring a large mass term, with one-loop quadratic corrections
being cancelled by new gauge bosons and  partners of the top quark. A simple
implementation of the Little Higgs concept with a single global symmetry group is the
\emph{Littlest} Higgs model \cite{Littlest}. However, owing to tree-level
contributions of the new particles to the oblique electroweak parameters,
electroweak precision data requires $f$ to be above 5 TeV \cite{lhew}. On the
other hand a scale as low as 1 TeV is required to avoid fine-tuning of the Higgs
mass.

This problem can be circumvented by imposing  a discrete symmetry, called
\emph{T-parity} \cite{LHT, wudka}. Under this symmetry, the Standard Model (SM)
fields are T-even, while the new TeV-scale particles are odd, effectively
forbidding all tree-level interactions between one of the new heavy degrees of
freedom and SM particles. Therefore, the new particles can only be generated in
pairs, which is reminiscent of R-parity in supersymmetric theories. Besides
satisfying the electroweak constraints even for $f < 1$ TeV, an exactly realized
T-parity also leads to the lightest T-odd particle being stable and, if neutral,
a good candidate for (cold) dark matter.

However, it was pointed out by Hill and Hill \cite{hillsq1,hillsq2} that typical
models of strongly interacting symmetry breaking would lead to a
Wess-Zumino-Witten (WZW) term \cite{wzw} which is odd under T-parity\footnote{It
is possible to construct models which do not have WZW terms or where T-parity is
not broken by these terms \cite{Krohn:2008ye,Csaki:2008se}. This avenue
will not be explored further in this paper.}. The structure of the WZW term can
be derived from topological considerations and depends only on the pattern of
the global and gauged symmetry groups. The breaking of T-parity by the WZW term,
though suppressed by the large symmetry breaking scale, rules out the lightest
T-odd particle as a dark matter candidate, since this particle would decay
promptly into gauge bosons \cite{Barger:2007df}. Nevertheless, if the WZW term
is a priori the only source of T-parity breaking, the electroweak precision
constraints could still be satisfied.

In this paper, we analyze the effect of the WZW term in the Littlest Higgs model
with T-parity (LHT) further.  The relevant interactions induced
by this
term are derived, and their gauge invariance is shown. Furthermore, it is
demonstrated that the WZW term cannot be the only T-parity violating operator,
but that other T-odd terms are needed in the Lagrangian to make the theory
consistent. Equipped with these results, we discuss the constraints on the model
from LEP and Tevatron data and comment on the surprisingly rich phenomenological prospects for LHC.

After reviewing the LHT model and specifying the notations in section
\ref{sc:model}, the T-odd interactions induced by the WZW term are studied in
section \ref{sc:wzw}. In section \ref{sc:pheno}  the T-parity violating signals
at LEP and hadron colliders are investigated. Finally, conclusions are given
in section \ref{sc:concl}.


\section{The Littlest Higgs model with T-parity}
\label{sc:model}

Here the main aspects of the model are reviewed, following the detailed
description in Refs.~\cite{hubmed,Blanke:2006eb}.

The Littlest Higgs model is  based on a SU(5)/SO(5) symmetry breaking pattern. A global SU(5) symmetry is broken down to SO(5) by a vacuum expectation value of the form
\begin{equation}
\langle \Sigma \rangle = \Sigma_0 = \begin{pmatrix}
&&&1&\\&&&&1\\&&1&&\\1&&&&\\&1&&&
\end{pmatrix}
\end{equation}
for a field $\Sigma$ transforming in the two-index symmetric representation of SU(5). The generators of SU(5) are split up into a set of ten unbroken generators $T^a$ that generate the unbroken SO(5) subgroup and a set of 14 broken generators $X^a$. 

The Goldstone modes of the broken generators are
implemented in a non-linear sigma model with a breaking scale $f$,
\begin{equation}
{\cal L}_\Sigma = \frac{f^2}{4} \text{Tr} |D_\mu \Sigma|^2,\label{lag0}
\end{equation}
with
\begin{equation}
\Sigma = e^{i \Pi/f} \Sigma_0 e^{i \Pi^\top/f} = e^{2i \Pi/f} \Sigma_0,
\end{equation}
where  $\Pi = \pi_a X^a$ is the Goldstone matrix. 
A $[\mathrm{SU(2)} \times \mathrm{U(1)}]^2$ subgroup of SU(5) is gauged \cite{Littlest}, with
associated gauge bosons $W^a_{1,2}$ and $B_{1,2}$, respectively.
In terms of $2\times 2$, $1\times 1$ and $2\times 2$ blocks, the gauge group generators
are given by
\begin{align}
Q_1^a &= \begin{pmatrix}
\sigma^a/2 & 0 & 0 \\
0 & 0 & 0 \\
0 & 0 & 0 
\end{pmatrix}, 
&
Y_1 &= \frac{1}{10} \begin{pmatrix}
3 & 0 & 0 \\
0 & -2 & 0 \\
0 & 0 & -2 
\end{pmatrix},
\\
Q_2^a &= \begin{pmatrix}
0 & 0 & 0 \\
0 & 0 & 0 \\
0 & 0 & -\sigma^{a*}/2 
\end{pmatrix},
&
Y_2 &= \frac{1}{10} \begin{pmatrix}
2 & 0 & 0 \\
0 & 2 & 0 \\
0 & 0 & -3 
\end{pmatrix},
\end{align}
where $\sigma^a$, $a=1,2,3$ are the Pauli matrices. The covariant derivative reads
\begin{equation}
D_\mu\Sigma \equiv \partial_\mu\Sigma -
\sum_{k=1,2} \bigl [ g_k W_{k,\mu}^a (Q_k^a\Sigma + \Sigma Q_k^{aT}) +
g'_k B_{k,\mu} (Y_k \Sigma + \Sigma Y_k) \bigr ].
\end{equation}
The vacuum $\Sigma_0$ breaks the gauge symmetry $[\mathrm{SU(2)} \times \mathrm{U(1)}]^2$ down to
the diagonal subgroup, giving one set of gauge bosons with masses
of order $f$, while the
other set remains massless at this stage and is identified with the Standard Model gauge
bosons. The Goldstone matrix $\Pi$ is explicitly given by
\begin{equation}
\Pi = \begin{pmatrix}
\omega/2 -\eta  /\sqrt{20}\,  \Eins & H/\sqrt{2} & \Phi \\
H^\dagger/\sqrt{2} & \sqrt{4/5}\eta & H^\top/\sqrt{2} \\
\Phi^\dagger & H^*/\sqrt{2} & \omega^\dagger/2 - \eta/\sqrt{20}\, \Eins
\end{pmatrix},
\end{equation}
where $H$ is the Little Higgs doublet, $\Phi$ is a complex triplet under
$(\mathrm{SU(2)}\times \mathrm{U(1)})_{SM}$, which receives a mass of ${\cal O}(f)$, and the real triplet field $\omega = \omega^a \sigma^a$ and the real singlet $\eta$ are eaten by
the heavy gauge bosons ($\Eins$ is the $2\times 2$ identity matrix).

The Littlest Higgs model can be supplemented by a discrete $Z_2$ symmetry called
T-parity \cite{LHT}, with SM particles being even ($T=+1$), and non-SM
particles odd ($T=-1$) under this symmetry. Their couplings to 
the non-linear sigma fields 
generate masses of order $f$ for the T-odd particles.
In the gauge sector, T-parity is realized by the automorphism $T^a \to T^a$ and
$X^a \to -X^a$. As a result, T-parity interchanges the two sets of gauge bosons,
\begin{equation}
W^a_1 \leftrightarrow W^a_2, \qquad
B_1 \leftrightarrow B_2.
\end{equation}
T-parity requires the two sets of gauge couplings to be
identical: $g_1=g_2=\sqrt{2}g$ and $g'_1=g'_2=\sqrt{2}g'$.
The gauge bosons form a light and a heavy linear combination:
\begin{align}
W_L^a &= \tfrac{1}{\sqrt{2}}(W_1^a + W_2^a), && \text{(T-even)} \\
B_L &= \tfrac{1}{\sqrt{2}}(B_1 + B_2),
\intertext{with masses from usual electroweak symmetry breaking, and}
W_H^a &= \tfrac{1}{\sqrt{2}}(W_1^a - W_2^a), && \text{(T-odd)} \\
B_H &= \tfrac{1}{\sqrt{2}}(B_1 - B_2),
\end{align}
with masses of order $f$ generated from the kinetic term of the non-linear sigma
model. After electroweak symmetry breaking, the light gauge bosons mix to form
the usual physical states of the SM, $A_L = \cw B_L - \sw W_L^3$, $Z_L = \sw B_L
+ \cw W_L^3$ and $W_L^\pm = (W_L^1 \mp W_L^2)/\sqrt{2}$. Here, as usual,
$\sw$ and  $\cw$ denote the sine and cosine of the weak mixing angle. Similarly, a small
mixing of order ${\cal O}(v^2/f^2)$ is introduced between $B_H$ and $W_H^3$
through electroweak symmetry breaking, yielding
\begin{align}
A_H &= \cos\theta_H \, B_H - \sin\theta_H W_H^3, &
  M^2_{A_H} &= \frac{g'^2}{5}f^2 - \frac{g'^2}{4} v^2 + 
  {\cal O}\Bigl (\frac{v^4}{f^2}\Bigr ),
\\
Z_H &= \sin\theta_H \, B_H  + \cos\theta_H W_H^3, &
  M^2_{Z_H} &= g^2 f^2 - \frac{g^2}{4} v^2 + 
  {\cal O}\Bigl (\frac{v^4}{f^2}\Bigr ),
\\
&&
\sin\theta_H &= \frac{gg'}{4g^2 - \frac{4}{5} g'^2} \, \frac{v^2}{f^2}
  + {\cal O}\Bigl (\frac{v^4}{f^4}\Bigr ),
\intertext{and}
W_H^\pm &= (W_H^1 \mp W_H^2)/\sqrt{2}, &
  M^2_{W^\pm_H} &= g^2 f^2 - \frac{g^2}{4} v^2.
\end{align}
The $A_H$ will be referred to as heavy
photon throughout this text. It is always lighter than the other T-odd gauge
bosons and thus a good candidate for the LTP (lightest T-odd particle) and dark
matter, if T-parity is an exact symmetry. Note  that
the mixing between the heavy photon and $Z_H$,  is numerically small
and leads to corrections at the 1\% level at most.

In the scalar sector, T parity is defined as
\begin{equation}
\Pi \to - \Omega \Pi \Omega, \qquad
\Omega = {\rm diag}(1,1,-1,1,1),
\end{equation}
such that $H$ is T-even while $\Phi$, $\omega$ and $\eta$ are T-odd.

The kinetic term  (\ref{lag0}) is not the full non-linear sigma model  Lagrangian but  just the first term in an expansion in external momenta $p$. The  higher order terms that have to be added to (\ref{lag0}) to
cancel divergencies that appear in perturbation theory  are suppressed by powers of $(p/\Lambda)$, where $\Lambda$ is the intrinsic cutoff of the theory beyond which ordinary perturbation theory breaks down. 

In Little Higgs models $\Lambda = 4\pi f$ is typically of the order of $10$~TeV, and the
phenomenology at the TeV scale is well described by (\ref{lag0}). Exceptions are
possible if the
lowest order Lagrangian possesses more symmetries than the full model. In that case higher order
terms have to be taken into account and may change the phenomenology significantly.

T-parity also requires a doubling of the left-chiral fermion sector.
Each left-handed T-even (SM) fermion is accompanied by a T-odd partner $f_H$ 
(mirror fermion) with mass \cite{mirrormass}
\begin{equation}
m_{f_{H,i}} = \sqrt{2} \kappa_{i} f 
 + {\cal O}\Bigl (\frac{v^2}{f}\Bigr ),
\end{equation}
where the Yukawa couplings $\kappa_{i}$ 
can in general depend on the fermion species
$i$. 

The implementation of the mass terms for the mirror fermions also introduces
T-odd
SU(2)-singlet fermions, which
may receive large masses and do not mix with the SU(2)-doublets $f_H$.
Here it is therefore assumed that these extra singlet fermions are too heavy to
be observable at current or next-generation collider experiments.

The top sector requires an additional T-even fermion $t'_+$ and one T-odd
fermion $t'_-$ to cancel quadratic divergencies to the Higgs mass. Both
particles obtain order $f$ masses. We will not discuss the top sector of the
Littlest Higgs model here, but refer the reader to
Refs.~\cite{hubmed,Blanke:2006eb} for further details. 
The Feynman rules of the Littlest Higgs model with T-parity are summarized in
Ref.~\cite{Blanke:2006eb}.


\section{The WZW term in the Littlest Higgs model}
\label{sc:wzw}

\subsection{The Wess Zumino Witten term}

The nontrivial vacuum structure of the
Littlest Higgs leads  one to include the Wess Zumino Witten term \cite{wzw} in
the effective Lagrangian  \cite{hillsq1}. It consists of two parts,
\begin{align}
        \Gamma_{WZW} & =\frac{N}{48\pi^2}\left( \Gamma_0(\Sigma) + 
	 \Gamma (\Sigma,A_l,A_r)\right) .
\end{align}
Here $\Gamma_0$ is the ungauged WZW term that can be expressed as integral over
a five-dimensional manifold with spacetime as its boundary\cite{wzw}, whereas
$\Gamma (\Sigma,A_l,A_r)$ is the gauged part of the WZW action that can be
written as an ordinary four-dimensional spacetime integral. The explicit form of
$\Gamma(\Sigma,A_l,A_r)$ and a prescription how to relate the gauge fields
$A_l$, $A_r$ to those appearing in the Littlest Higgs are given in
Ref.~\cite{hillsq1}: 
\begin{equation}
A_{l,r} = \sqrt{2} \bigl [ g (W_L^a Q_L^a \mp W_H^a Q_H^a) +
		g' (B_L Y_L \mp B_H Y_H) \bigr ].
\end{equation}

The Integer $N$ depends on the UV completion of the Littlest Higgs
model. In strongly coupled UV-completions, where the Little Higgs is a composite
particle of some underlying Ultracolor theory \cite{Katz:2003sn}, $N$ will
equal the number of ultracolors, $N=N_{uc}$. 

The WZW term is T-odd by construction, i.e. it changes sign under a T-parity transformation. The fact that  $\Gamma_{WZW}$ violates T-parity and that its coefficient $N$ can not be chosen arbitrarily make the WZW term stand out from other higher order terms in the expansion of the non-linear sigma model lagrangian.

In those cases where the UV completion demands a nonzero $N$, T-parity cannot be a fundamental symmetry of the full theory, instead it has to be seen as accidental symmetry of the lowest order effective
Lagrangian.  This is the point of view we want to adopt in this work. 

\subsection{Gauge invariance}

The WZW term is not manifestly gauge invariant, rather under a gauge
transformation
\begin{align}
        \Sigma \rightarrow e^{i\eps_l} \Sigma e^{-i\eps_r}, \qquad 
	A_l^\mu \to A_l^\mu + \partial^\mu \eps_l +i [\eps_l, A_l^\mu],\qquad 
	A_r^\mu \to A_r^\mu + \partial^\mu \eps_r +i [\eps_r, A_r^\mu],
\end{align}
 it transforms as $\Gamma_{WZW} \rightarrow \Gamma_{WZW} + \delta\Gamma_{WZW}$, with $\delta \Gamma_{WZW}$ given by
\begin{align}
        \delta \Gamma_{WZW}= -\frac{N}{24\pi^2}\int d^4x \,
	\epsilon_{\mu\nu\rho\sigma} \text{Tr}\left[\epsilon_l 
	\left( \partial^\mu A_l^\nu \partial^\rho A_l^\sigma -
	{\textstyle \frac{i}{2}} \partial^\mu (A_l^\nu A_l^\rho A_l^\sigma )
	\right) - (L\rightarrow R) \right],\label{anomaly}
\end{align}
reproducing the well-known nonabelian chiral anomaly  \cite{wzw}. 
In order to restore gauge invariance, a sector must be added to the theory whose gauge variation cancels (\ref{anomaly}) exactly. Various options to cancel the anomaly are 
discussed in  \cite{hillsq2}. 

One possible way to to cancel the anomaly directly at the level
of the underlying ultracolor theory is by introducing a set of spectator leptons
with $U(1)_1$ and $U(1)_2$ charges chosen such that they directly cancel the
anomalies from the ultrafermions.
Making the spectator leptons sufficiently heavy  allows one to
neglect their contributions to physical observables, without affecting the
anomaly cancellation. 

The anomalous couplings in $\Gamma_{WZW}$ are the terms with three or four gauge bosons, with an odd number of T-odd gauge bosons. For example the three gauge boson terms with one T-odd gauge boson are of the form
$\eps_{\mu\nu\rho\sigma} V_H^\mu V^\nu \partial^\rho V^\sigma$, where $V_H$ is
any T-odd gauge boson and $V$ denote SM gauge bosons.
Independent of the actual implementation, any anomaly canceling sector does at least cancel all these terms. 

There could be additional effects from the anomaly canceling sector that do depend on the details of its implementation. We will here assume that these effects  can be decoupled, as in the example above, or at least are suppressed by some sufficiently large scale, and leave the details to further studies. 

\

With the anomalous couplings cancelled, the leading T-odd interactions now appear at order $(1/f^2)$ in the expansion of $\Gamma_{WZW}$. For example three gauge boson interactions with two SM gauge bosons and one T-odd gauge boson are generated by  $\eps_{\mu\nu\rho\sigma}
H^\dagger H/f^2 V_H^\mu V^\nu \partial^\rho V^\sigma$ once electroweak symmetry is broken. 

The systematic expansion of $\Gamma_{WZW}$  leads to a large number of T-parity violating interactions. To leading order in $(1/f)$ the part of the WZW term containing one neutral T-odd gauge boson is given by
\begin{align}
        \Gamma_n &= \frac{Ng^2 g'}{48 \pi^2 f^2}\int d^4x \, (v+h)^2
\epsilon_{\mu\nu\rho\sigma}\times \label{WZW1} \\
        & \quad \left[ - {\textstyle\frac{6}{5}} A_H^\mu \left( c_w^{-2} Z^\nu
\partial ^\rho Z^\sigma +W^{+\nu} D_A^\rho W^{-\sigma} + W^{-\nu}D_A^\rho
W^{+\sigma} +i (3 g c_w + g' s_w) W^{+\nu}W^{-\rho}Z^\sigma \right)+ \right.
\nonumber \\
        & \left. \quad t_w^{-1} Z_H^\mu \left(2 c_w^{-2}  Z^\nu \partial^\rho
Z^\sigma +W^{+\nu}D_A^\rho W^{-\sigma} + W^{-\nu}D_A^\rho W^{+\sigma} -2 i(2 g
c_w +g's_w)W^{+\nu}W^{-\rho}Z^\sigma \right) \right] \nonumber
\end{align}
while the part containing one charged T-odd gauge boson reads
\begin{align}
        \Gamma_c &= \frac{Ng^2 g'}{48 \pi^2 f^2}\int d^4x \, (v+h)^2
\epsilon_{\mu\nu\rho\sigma}\times \nonumber \\
        & \quad \left[ 2 W_H^{+\mu} W^{-\nu}(-c_w \partial^\rho A^\sigma +s_w
\partial^\rho Z^\sigma) +c_w  W_H^{+\mu} D_A^\nu W^{-\rho}(-A^\sigma +(2 t_w
+t_w^{-1})Z^\sigma) +\right. \nonumber\\
        &\quad \left.   c_w D_A^\mu W_H^{+\nu} W^{-\rho} (A^\sigma +t_w^{-1}
Z^\sigma)\right] + h.c. \label{WZW2},
\end{align}
written in unitary gauge. The vacuum expectation value of the Higgs field is denoted by $v$, $h$ is the physical Higgs boson and we defined $D_A^\mu W^{\pm\nu}=(\partial^\mu \mp i e A^\mu)W^{\pm \nu}$. Furthermore $s_w$, $c_w$ and $t_w$ denote the sine, cosine and tangent of the weak mixing  angle, respectively. 

We do not write other parts of the WZW term here, instead all T-violating
vertices with up to four legs have been tabulated in appendix
\ref{sec:feynrules}, including the interactions of the complex
triplet $\Phi$.
These Feynman rules have further been implemented into a model file for
\textsc{CalcHEP 2.5} \cite{calchep,modelfile}.

Because of (\ref{WZW1}) the heavy photon can decay either into a pair of
$Z$-bosons or into a $W^+W^-$ pair, with a decay width of order ${\cal O} ( \mathrm{eV})$ \cite{Barger:2007df}. This clearly rules out the $A_H$ as dark
matter candidate. A more detailed analysis, in particular for the case where the
decay into real SM gauge bosons is kinematically forbidden, will be performed in
section \ref{sc:pheno}. 

\vspace{1em}
 
The gauge invariance of the WZW term can be verified using 
Ward identities for the three-point functions
involving massive gauge bosons. These identities can be derived in a similar
way as the
Ward identities for three-boson vertices in the SM \cite{Denner:1994nn}.
For example vertices involving the heavy photon $A_H$ have to satisfy
\begin{align}
        k^\nu_2 \Gamma_{\mu\nu\rho}^{A_H Z Z}(k_1,k_2,k_3) - i m_Z
\Gamma_{\mu\rho}^{A_H G^0 Z}(k_1,k_2,k_3)=0, \label{wi1} \\
        k^\nu_2 \Gamma_{\mu\nu\rho}^{A_H W^+W^-}(k_1,k_2,k_3)-m_W
\Gamma_{\mu\rho}^{A_H G^+ W^-}(k_1,k_2,k_3)=0 . \label{wi2}
\end{align}
At the tree level these identities have simple interpretations in terms of
Feynman graphs, as shown in Fig.~\ref{fig:wi1}. 
\begin{figure}[tb]
\begin{center}
\begin{picture}(400,100)(0,0)
        {\SetWidth{1.5}
        \Photon(60,50)(120,50){3}{5}}
        \Photon(120,50)(180,80){3}{6}
        \Photon(120,50)(180,20){3}{6}
        \Vertex(120,50){2}
        {
        \Text(52,50)[r]{$k_2^\nu\; \cdot $}}
        \Text(90,60)[b]{$A_{H,\mu}$, $k_1$}
        \Text(150,75)[b]{$W^+_\nu$, $k_2$}
        \Text(150,25)[t]{$W^-_\rho$, $k_3$}
        {\Text(222,50)[r]{$-\;m_W \;\cdot$} }
        {\SetWidth{1.5}
        \Photon(230,50)(290,50){3}{5}}
        \Vertex(290,50){2}
        \DashLine(350,80)(290,50){3}
        \Photon(290,50)(350,20){3}{6}
        \Text(260,60)[b]{$A_{H,\mu}$, $k_1$}
        \Text(320,75)[b]{$G^+$, $k_2$}
        \Text(320,25)[t]{$W^-_\rho$, $k_3$}
        { \Text(360,50)[l]{$= $ 0}}
\end{picture}
\end{center}
\vspace{-8mm}
\mycaption{Tree level Ward identity for the $A_H W^+W^-$ vertex, all momenta
incoming}\label{fig:wi1}
\end{figure}
%
%
%
Using the gauge boson-Goldstone boson vertices of Tab.~\ref{tab:gold} we have
checked explicitly that $[SU(2)\times U(1)]_{SM}$ gauge invariance is respected
by all interactions coming from equations (\ref{WZW1}) and (\ref{WZW2}).

\subsection{Divergences and counterterms}

Apart from the tree level interactions additional T-parity violating processes
are induced at the loop level. These are especially important when corresponding
tree level processes are kinematically forbidden. In particular,  when $M_{A_H}
< 2 M_W$, the heavy photon cannot decay into real SM gauge bosons, and decays
induced by one-loop processes have to be taken into account. 
\begin{figure}[tb]
\vspace{3mm}
\begin{center}
\psfig{figure=AHFF.epsi, width=5cm}\hspace{.5cm}
\psfig{figure=AHFF2.epsi, width=5cm}\hspace{.5cm}
\psfig{figure=AHFF3.epsi, width=5cm}
\\[.5em]
\anc\hspace{1.5cm}(a)\hspace{5cm}(b)\hspace{4.7cm}(c)\\[.5em]
\psfig{figure=AHFF4.epsi, width=5cm}\hspace{.5cm}
\psfig{figure=AHFF5.epsi, width=5cm}\hspace{.5cm}
\psfig{figure=AHFF6.epsi, width=5cm} \\[.5em]
\anc\hspace{1.5cm}(d)\hspace{5cm}(e)\hspace{4.7cm}(f)
\end{center}
\vspace{-5mm}
\mycaption{Loop induced decay of $A_H$ into SM quarks/leptons. Thick
lines indicate T-odd propagators. $q=(u,d,c,s,b)$, $\tilde{q}=(d,u,s,c,t)$, and
similar for $l$, $\tilde{l}$. }
\label{fig:ahff}
\end{figure}

The most important processes are of the type shown in
Fig.~\ref{fig:ahff}, where the heavy photon couples to two light T-even fermions
via a triangle loop. A similar set of graphs also couple $Z_H$ and $W_H^{\pm}$
to SM fermions.  Since the three-boson vertex involves one power of
the loop momentum, graphs of this type are logarithmically divergent. 

The counterterms needed to cancel these divergencies are of the form
\begin{align}
        \mathcal{L}_{ct} &= \bar{f} \gamma_\mu 
        \left( c_L^f P_L + c_R^f P_R \right) f A_H^\mu, \label{ct1} \\
        c_i^f &= c_{i,\epsilon}^f \left( \frac{1}{\epsilon} + \log \mu^2 +
        {\cal O}(1) \right).
\end{align}
The coefficients $c_i(\mu)$ of the counterterms are determined as follows. The
scale dependence of the above loop processes must be cancelled by the scale
dependence of the $c_i(\mu)$. Naturalness arguments then suggest that an
${\cal O}(1)$
change in the renormalization scale should be compensated by an ${\cal O}(1)$
change in
the $c_i(\mu)$. Therefore these coefficients are given, up to ${\cal O}(1)$
factors, by
the coefficients of the leading $1/\epsilon$ divergence in dimensional
regularization of the above loop diagrams. 
The resulting coefficients are given in Tab.~\ref{tab:ct1}. 
Since the $A_H$ only couples very weakly to fermions, the contributions of
diagrams (e) and (f) in Fig.~\ref{fig:ahff} have been neglected. An alternative, gauge invariant formulation of the counterterms (\ref{ct1}) is discussed in appendix \ref{sec:ct}.
%
\begin{table}[tp]
\renewcommand{\arraystretch}{1.5}
\begin{center}
\begin{tabular}{|l|c|c|}\hline
        Particles & $c_{L,\epsilon}^f$ & $c_{R,\epsilon}^f$  \\ \hline \hline 
        $A_H e^+e^-$ 
        &$ \textstyle \frac{9 \hat{N}}{160 \pi^2}\frac{v^2}{f^2} g^4 g'
\left(4+(c_w^{-2}-2t_w^2)^2\right)$  
        & $\textstyle -\frac{9 \hat{N}}{40 \pi^2} \frac{v^2}{f^2}g'^5  $  \\ \hline 
        $A_H \bar{\nu}\nu  $ 
        &  $ \textstyle \frac{9 \hat{N}}{160 \pi^2}\frac{v^2}{f^2} g^4 g'
\left(4+c_w^{-4} \right)$
        & 0 \\ \hline 
        $ A_H \bar{u}_a u_b$ 
        & $ \textstyle - \frac{\hat{N}}{160 \pi^2}\frac{v^2}{f^2} g^4 g' \left( 36 +(3
c_w^{-2}-4 t_w^2 )^2\right) 
	  \delta_{ab}$
        & $ \textstyle - \frac{\hat{N}}{10 \pi^2}\frac{v^2}{f^2} g'^5 \delta_{ab}$ \\
\hline
        $ A_H \bar{d}_a d_b$ 
        & $ \textstyle - \frac{\hat{N}}{160 \pi^2}\frac{v^2}{f^2} g^4 g' \left( 36 +(3
c_w^{-2}-2 t_w^2 )^2\right)  
	  \delta_{ab}$
        & $ \textstyle - \frac{\hat{N}}{40 \pi^2}\frac{v^2}{f^2} g'^5 \delta_{ab}$  \\
\hline
\end{tabular}
\mycaption{Coefficients for the counterterm (\ref{ct1}). Here $\textstyle \hat{N}=
\frac{N}{48\pi^2}$ denotes the coefficient of the WZW term, while $a,b$ indicate
the color indices of the external quarks.}
\label{tab:ct1}
\end{center}\vspace*{-\baselineskip}
\end{table}
%

Another important set of diagrams arises from those in Fig.~\ref{fig:ahff} by
replacing the $A_H$ with a $Z$ boson and one of the fermions with its mirror
partner. These diagrams, as well as the corresponding diagrams where the $Z$ is
replaced by $W^{\pm}$, are again logarithmically divergent and require
T-violating counterterms.\footnote{The loop that couples the photon to a
fermion-mirror fermion pair is finite, as required by gauge invariance.} These
may become important in scenarios where one of the mirror fermions is the LTP.

Other T-violating counterterms induced at the one loop level are not relevant
for phenomenology for almost all reasonable choices of parameters in the
Littlest Higgs model. 

Mixing between T-even and T-odd gauge bosons is also induced by loop diagrams and may affect electroweak precision observables. The existing one loop graphs vanish due to the
antisymmetry of the $\epsilon$-tensor, so the first contributions come only at the two loop level. 
 The nonvanishing two loop diagrams are shown in Fig.~\ref{fig:ahz}. 
\begin{figure}[tb]
\begin{center}
\psfig{figure=AHZ1.epsi, width=5cm}\hspace{.5cm}
\psfig{figure=AHZ2.epsi, width=5cm}\hspace{.5cm}
\psfig{figure=AHZ3.epsi, width=5cm}
\\[.5em]
\anc(a)\hspace{5cm}(b)\hspace{5cm}(c)\\[.5em]
\psfig{figure=AHZ4.epsi, width=5cm}\hspace{.5cm}
\psfig{figure=AHZ5.epsi, width=5cm}\hspace{.5cm}
\psfig{figure=AHZ6.epsi, width=5cm} \\[.5em]
\anc(d)\hspace{5cm}(e)\hspace{5cm}(f)
\end{center}
\vspace{-5mm}
\mycaption{Two loop diagrams that contribute to  $A_H$-$Z$ boson mixing. As
above thick lines indicate T-odd propagators.  All types of fermions and
mirror-fermions are allowed. }
\label{fig:ahz}
\end{figure}

The relevant counterterms are of the form
\begin{align}
        \mathcal{L}_{ct} &= \frac{c^{V_LV_H}}{4} 
	 (\partial_\mu V_{L,\nu} - \partial_\nu V_{L,\mu})
	 (\partial^\mu V_H^\nu - \partial^\nu V_H^\mu), \label{ct2} \\
	c^{V_LV_H} &= c_{\epsilon}^{V_LV_H} \left( \frac{1}{\epsilon} + \log \mu^2 +
	{\cal O}(1) \right),
\end{align}
where $V_L \in \{A_L, Z_L, W^\pm_L\}$ and $V_H \in \{A_H, Z_H, W^\pm_H\}$.
For the gauge boson mixing terms, the leading $1/\epsilon$ divergence is not completely 
determined in the LHT model. The reason is that  the LHT model as a
low-energy effective theory has an ``incomplete'' fermion content whose 
 [U(1)$_i \times $SU(2)$_j \times $SU(2)$_j$]
gauge anomalies ($i,j=1,2$) must be cancelled (see above) by an interacting UV completion.
If not specified, the  ${\cal O}(1)$
uncertainty remains for the $1/\epsilon$ coefficients of the T-violating gauge
boson mixing counterterms.
 For this reason we only list the parametric
dependence of the coefficients $c_{\epsilon}^{V_LV_H}$ with an undetermined
prefactor which is expected to be close to one:
\begin{align}
c_{\epsilon}^{V_LV_H} &= {\rm const.} \times \frac{\hat{N}}{(4\pi)^4} \frac{v^2}{f^2}
\begin{aligned}[t]
g^5 g' \Bigl (&B_0(k^2,0,0) + 2 \tfrac{m_t^2}{k^2} \, B_0(0,m_t^2,m_t^2) \\
&- (1+2\tfrac{m_t^2}{k^2}) \, B_0(k^2,m_t^2,m_t^2) \Bigr ),
& \quad & V_i = A_i,Z_i,
\end{aligned}
\\
c_{\epsilon}^{W_L^\pm W_H^\pm} &= {\rm const.} \times 
\frac{\hat{N}}{(4\pi)^4} \frac{v^2}{f^2}
\begin{aligned}[t]
g^5 g' \Bigl (&B_0(k^2,0,0) + 
\tfrac{m_t^2}{k^2}(1-2\tfrac{m_t^2}{k^2}) \, B_0(0,0,m_t^2) \\
&-\tfrac{(k^2-m_t^2)(k^2+2m_t^2)}{k^4} \, B_0(k^2,0,m_t^2) - \tfrac{m_t^2}{k^2} 
\Bigr ),
\end{aligned}
\end{align}
where all SM Yukawa couplings except the top Yukawa coupling have been set to
zero, $\hat{N} = \frac{N}{48\pi^2}$ and ``const.'' stands for a complex ${\cal O}(1)$ constant, which depends on
$V_L$ and $V_H$. Here $B_0$ is the
usual standard one-loop self-energy function and $k$ is the external gauge boson
momentum. The mixing between T-even and T-odd gauge bosons induced by the WZW
term is very small due to the two-loop suppression and does not lead to
observable effects in electroweak precision observables. 
For $m_t \to 0$ the gauge boson mixing terms have to vanish owing to
gauge anomaly cancellation.

\section{Phenomenology of T-parity breaking effects}
\label{sc:pheno}


\subsection{Decays of $A_H$}
\label{ssc:ahdecay}

The leading decays of $A_H$ are induced by the $A_H W^+W^-$ and $A_H ZZ$ terms
in (\ref{WZW1}). For large enough $f$ the decay into real gauge bosons
is allowed and the corresponding partial widths are 

\begin{align}
        &\Gamma(A_H \rightarrow ZZ) = \frac{1}{2 \pi}\left( \frac{N
g'}{40\sqrt{3}\pi^2 }\right)^2 \frac{m_{A_H}^3 m_Z^2}{f^4} \left(1-\frac{4
m_Z^2}{m_{A_H}^2}\right)^{\frac{5}{2}}, \label{ahvv1}\\
        &\Gamma(A_H \rightarrow W^+W^-) = \frac{1}{ \pi}\left( \frac{N
g'}{40\sqrt{3}\pi^2 }\right)^2 \frac{m_{A_H}^3 m_W^2}{f^4} \left(1-\frac{4
m_W^2}{m_{A_H}^2}\right)^{\frac{5}{2}}.\label{ahvv2}
\end{align}
To leading order in $(1/f)$ this agrees with the result of Ref.~\cite{Barger:2007df}
if we set  $K=6\sqrt{5/3}$. 

The threshold for the decay into real gauge bosons, $m_{A_H}>2 m_W$, corresponds
to a value of $f=1070$~GeV. However previous studies have shown that values of
$f$ as low as $500$~GeV are consistent with electroweak precision data
\cite{hubmed,Hundi:2006rh}. In this region of parameter space, three-body
decays  via $A_H\rightarrow VV^*$ are dominant, where $V^*$ indicates an
off-shell SM gauge boson. 
For $f<600$~GeV the mass of $A_H$
even drops below $M_W$ and four-body decays via two virtual intermediate gauge
bosons have to be considered. 

Below $f\sim1200$~GeV the loop induced two body decays shown in
Fig.~\ref{fig:ahff} become relevant. 
A reliable estimate of the decay widths can be obtained by just using the finite,
scale independent part of the counterterms and setting to one the undetermined
${\cal O}(1)$  coefficient that enters the results through the
counterterms (\ref{ct1}) ( see section \ref{sc:wzw}).

The partial width of a massive gauge boson $V$ into a pair of
fermions with couplings
\begin{align}
        \mathcal{L}_{Vff} &= \bar{f} \gamma_\mu \left( \frac{r+l}{2} +
\frac{r-l}{2}\gamma_5 \right) f V^\mu, 
\end{align}
where $l$ and $r$ are the coefficients of the left and right chiral projectors, is given by
\begin{align}
        \Gamma_{V\rightarrow ff} & = \frac{N_{C_f} M_V}{48 \pi} \sqrt{1-\frac{4 m_f^2}{M_V^2}}
\left[\left(r-l \right)^2\left(1-\frac{4m_f^2}{M_V^2}\right) + \left(r+l\right)^2
\left(1+\frac{2m_f^2}{M_V^2}\right) \right],
\end{align}
$N_{C_f}$ denoting the number of colors of fermion species $f$. 

\begin{figure}
\psfig{figure=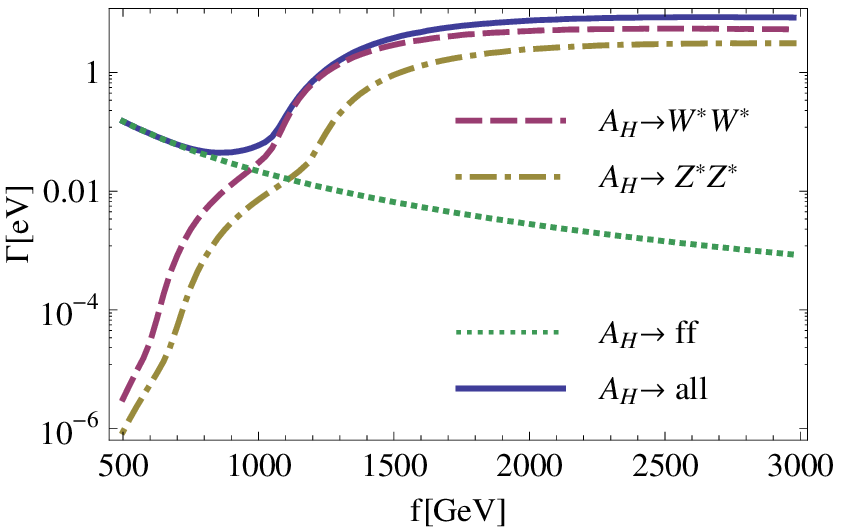, width=0.50\textwidth}\hfill%
{\psfig{figure=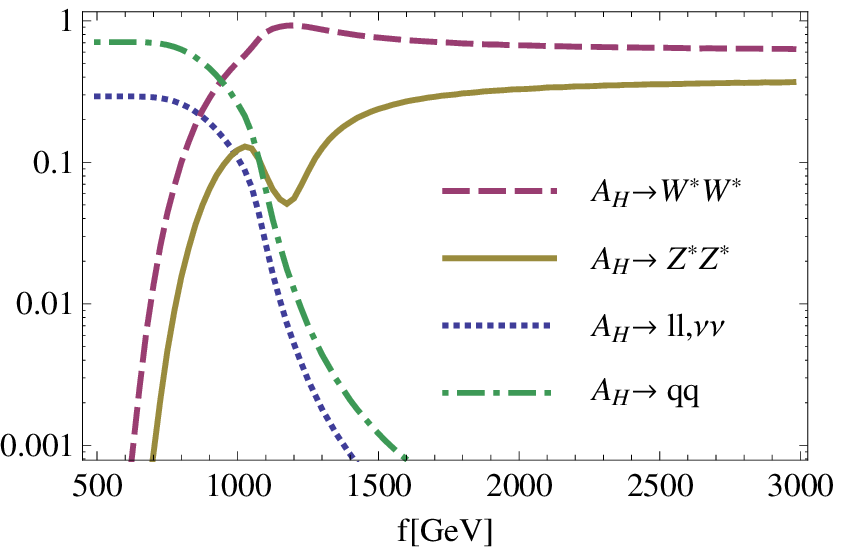, width=0.49\textwidth}}
\vspace{-2ex}
\mycaption{Left: Decay widths of $A_H$ into $Z^*Z^*$ (solid, red line), into $W^*W^*$ (dashed,
blue) and into fermion pairs (dotted, green) as well as the total width (thick black line),
for $N=3$. Right: Corresponding branching fractions. These are independent of $N$.}
\label{fig:ahwb}
\end{figure} 

The total width of $A_H$ for $N=3$, including the loop induced two body decays, is shown
in Fig.~\ref{fig:ahwb} together with the corresponding branching fractions.
Above $f=1500$~GeV, $A_H$ dominantly decays into two on-shell gauge bosons, with
a total width of $\Gamma_{A_H} \sim 1-2 \,\,\mathrm{eV}$. The fermionic decay
channels are negligible in this region.

Beneath the threshold for real $W$ production, $M_{A_H}<2m_W$,  the decay
phenomenology of $A_H$ changes dramatically. For $f$ below 
$\sim 1000$~GeV the decay into a light fermion pair becomes dominant; 
in approximately $10$\% of all cases, the decay is into a pair of charged
leptons with an invariant mass $m_{A_H}$ and an extremely small width of
 ${\cal O}$(eV). 
 
 The uncertainty in the fermionic widths due to the unknown ${\cal O}(1)$
coefficients in the counterterms (\ref{ct1}) may slightly change the value of $f$ where the fermionic decays become dominant,
however the overall picture does not change.


\subsection{Bounds from electroweak precision tests and direct detection at LEP}

T-parity in Little Higgs models evades the tension
between a low value of $f$ and electroweak precision tests (EWPT). Models
without T-parity are typically only  compatible with EWPT for
$f\geq 5$~TeV \cite{lhew}, while lower values of $f$ are favored by
naturalness.  

However, if T-parity is broken by the  WZW term, the situation is different and we do
not expect  disagreement with electroweak precision data, even for values of $f$
lower than $1$~TeV. One reason is that the coefficient of $\Gamma_{WZW}$,
$N/48 \pi^2$, is  very small for reasonable values of $N$. 
Furthermore, the T-odd operators affecting electroweak precision observables
are suppressed by loops, as discussed in section \ref{sc:wzw}, so that their
contribution is smaller than the experimental error of those observables.
We conclude that the Littlest Higgs model with anomalous T parity is not
constrained by electroweak precision data; in particular values of $f$ as low as
$500 \gev$ are  allowed. In addition the stringent bounds from dark matter
overproduction are evaded. 

\

While most of the T-odd particles are quite heavy, the $A_H$ is rather light and
could in principle have been produced and detected by the LEP experiments.
However,  the cross section for pair production of $A_H$ in
$e^+e^-$ collisions is smaller than $10^{-6}$~pb for all allowed values of $f$,
and thus invisible at LEP. T-violating single $A_H$ production is further
suppressed by $N/48 \pi^2$ and therefore also out of reach of the LEP experiments.


\subsection{Bounds from Tevatron}\label{ssc:teva}

The rates for pair production of heavy gauge bosons are relatively small also at
Tevatron. While the $A_H A_H$ production is suppressed due to its small
couplings, other combinations like $A_HZ_H$ or $W_H^+W_H^-$ are too
heavy to be produced in noticeable amounts. 

The situation is slightly different for the production of T-odd quark pairs. In
the LHT, their  mass is essentially a free parameter, only bound to lie between
 $100$~GeV and a few TeV, so they can in principle be light enough
to be produced in sizable amounts even at Tevatron.

The phenomenology of T-odd quarks  at Tevatron has been studied in Ref.~\cite{tevatron}, assuming a
common mass $m_{q_H}$ for the first two families of T-odd quarks and exact T-parity. It was found
that T-odd quarks are produced in sizeable numbers for $m_{q_H} <500$~GeV and are excluded for
$m_{q_H}<350$~GeV in the $2j +\Eslash_T$ channel.\footnote{For small T-odd masses  $m_{q_H}$
the dominant decay is $ q_H \rightarrow q A_H$ which yields a $j+\Eslash_T$ signal if
T-parity is unbroken.}

After including the WZW term the collider signatures of a T-odd quark change
completely. The main decay mode is still $q_H\rightarrow q A_H$, but the heavy
photon $A_H$ subsequently decays either into a pair of light fermions for small
$f$ or into a pair of (Standard Model) gauge bosons for larger values
of $f$. 

The cross section for the production of a $q_H \bar{q}_H$ pair in $p\bar{p}$ collisions
depends strongly on their mass $m_{q_H}$ and is nearly independent
of $f$. It is therefore sufficient to analyse the phenomenology for two
characteristic values of $f$, namely $f=750$~GeV where $A_H$ decays into fermion
pairs in more than 90\% of the cases, and $f=1500$~GeV where essentially all
$A_H$ decay into gauge boson pairs. The results of this section furthermore do
not depend on the actual value of $N$, as long as it is nonzero (see below). 

\begin{figure}\centering
\psfig{figure=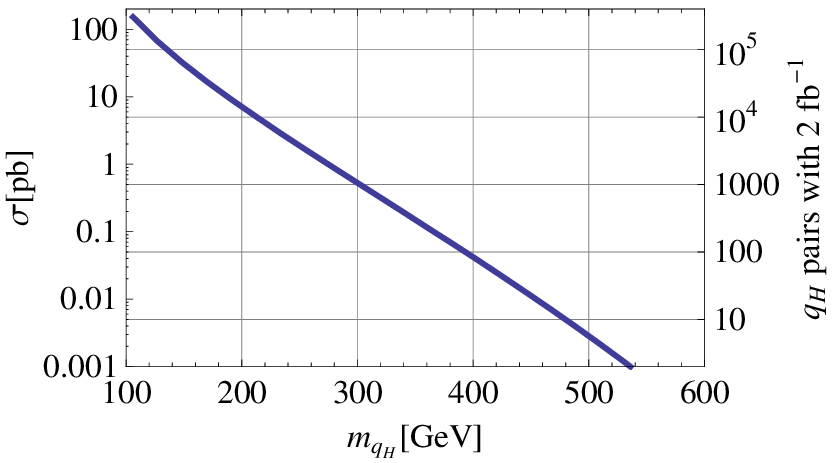, width=10cm}
\vspace{-1ex}
\mycaption{Cross section for first and second family mirror-quark pair production at Tevatron as a 
function of $m_{q_H}$. Right edge: Number of expected $q_H q_H$ pairs with $2$~fb$^{-1}$ of
integrated luminosity.}
\label{fig:teva1}
\end{figure} 

The cross section for the pair production of mirror quarks at Tevatron
is shown in Fig.~\ref{fig:teva1}, along with the number of
expected $q_H q_H$ pairs produced with $2$~fb$^{-1}$ of integrated luminosity, as
computed with {\sc CalcHEP}. The renormalization and factorization scales $\mu$ 
were chosen to be the invariant mass of the incoming partons. This is a
conservative choice as lower values of $\mu$ can increase the cross sections 
 by up
to 30\%. To reduce this scale dependence a full next-to-leading order
calculation of this process would be required. 

We will first consider the case $f=750$~GeV as  a representative value for
values of $f$ below $1000$~GeV, corresponding to $A_H$
masses of 80--150~GeV. Here we assume that $q_H$ is much heavier that $A_H$.
The case where $m_{q_H}$ is close to  $M_{A_H}$ is treated at the end of this section, while the
case where (some of)  the mirror fermions are lighter than $A_H$ is discussed in
section~\ref{f_ltop}. 

\begin{figure}\centering
\psfig{figure=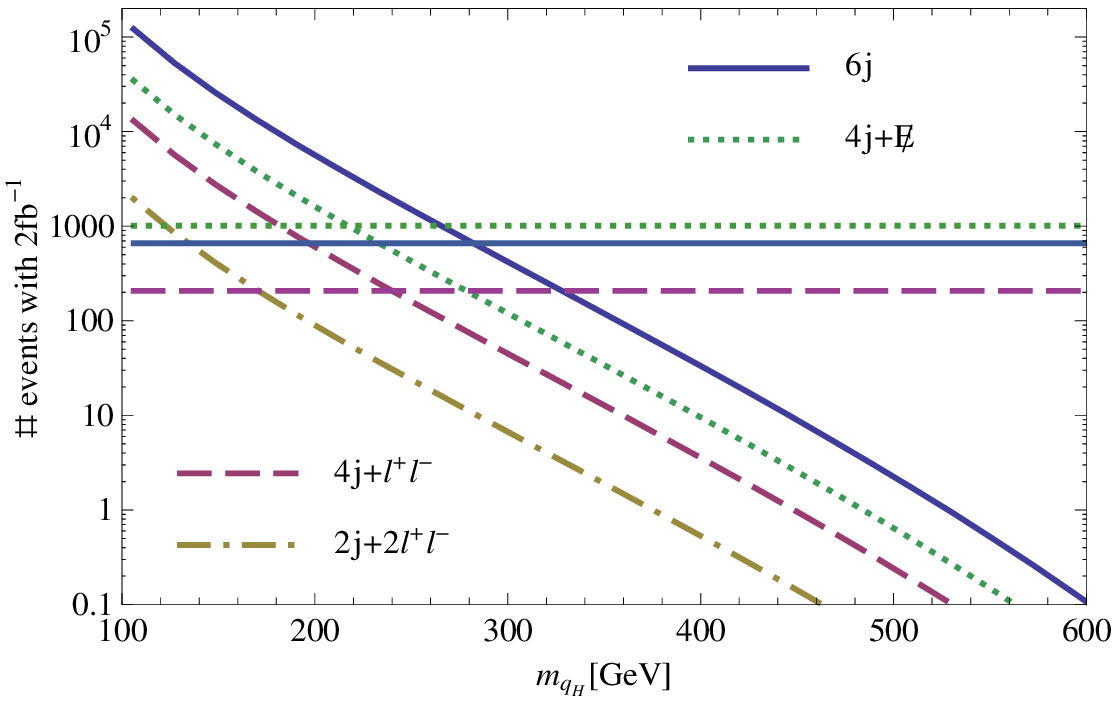, width=10cm}
\vspace{-2ex}
\mycaption{Event rates for the different decay channels of $q_Hq_H$ pairs for $f=750$~GeV with
$2$~fb$^{-1}$ at Tevatron, together with the corresponding SM backgrounds (horizontal
lines). No cuts have been applied to the signals. }
\label{fig:teva2}
\end{figure}

With $A_H$ decaying, there are various possible final states originating from a
$q_H\bar{q}_H$ pair. The highest rate results for the channel where both heavy photons
decay into quark pairs, leading to six jets. Further there are events with
four jets plus one lepton pair and events with two jets and two lepton pairs
with the same invariant mass. Finally also four jets plus missing energy, two
jets plus lepton pair plus missing energy and two jets plus missing energy are
possible, since one or both $A_H$'s can decay into neutrinos. Figure
\ref{fig:teva2} shows the expected number of events in the most promising
channels together with the corresponding SM backgrounds (as listed in
Ref.~\cite{cdf:vista}) for $2$ fb$^{-1}$, as a function of $m_{q_H}$.  Note that
we did not include detector acceptance effects in the computation of the signal,
 so that the experimentally observable rates could be slightly lower than the
numbers in the figure.

The most stringent bounds on $m_{q_H}$ could be derived from the six jet channel. Using the
preliminary data from the CDF Vista global search \cite{cdf:vista} at $2$~fb$^{-1}$ we find that
\begin{align}
        m_{q_H} > 350~\text{GeV}.
\end{align}
As mentioned, this bound could be improved by including NLO order corrections and appropriate cuts.

Due to their smaller background and cleaner signatures, the channels with leptonic final states
could in principle provide stronger bounds than the one derived from purely hadronic decays.
However in the interesting region above $m_{q_H} \gesim 300$~GeV their statistical significance
drops
quickly. A dedicated study of these final states still could yield more reliable the bounds, but
this is not possible with the Vista data alone. 

\

The analysis is more involved in the second case, $f>1000$~GeV, where $A_H$
dominantly decays into two gauge bosons. As can be seen in Fig.~\ref{fig:ahwb},
or from equations (\ref{ahvv1}) and (\ref{ahvv2}), the branching $B(A_H
\rightarrow W^+W^-)$ is larger than 60\% in that region, with a peak value of $\sim 95$\% around the
$W$-boson pair production threshold. 

The final states originating from the decay of a $q_H \bar{q}_H$ pair will consist of at
least ten particles. Events with eight quarks and two leptons are the most
common final state, followed by  ten jet events and events with six quarks and four leptons in the final state. In
most cases some or all of the leptons are neutrinos and escape detection, while
the fraction of events where all leptons come in charged pairs is rather
small. 

To give an idea of the range of $m_{q_H}$ that can be tested at Tevatron in this
case, in Tab.~\ref{tab:mqh} we list all final states with at
least one charged lepton together with the value of $m_{q_H}$ above which less
than 100 events are to be expected with $2$~fb$^{-1}$. 
%
\begin{table}[tp]
\renewcommand{\arraystretch}{1.5}
\begin{center}
\begin{tabular}{|c|c|c||c|c|c|}\hline
        Final State & BF~[\%] & $m_{q_H}$~[GeV] & Final State & BF~[\%] &  $m_{q_H}$~[GeV]  \\
\hline \hline
        $8j + l + \nu_l$ & 22.5 & 335 & $8j + l^+l^-$ & 4.4& 270  \\ \hline 
        $ 6j + 2\, (l \nu_l) $ & 13.9& 315 & $ 4j + 3\, l \nu_l $ & 4.1 & 265\\ \hline
        $ 6j + 2 \, (l^+l^-)$ & 0.4& 185 & $10 j$ & 21.9& 330 \\ \hline 
\end{tabular}
\mycaption{Possible final states from the decay of a $q_H\bar{q}_H$-pair for $f>1000$~GeV, together with
their branching fractions (second column) and the value of $m_{q_H}$ where the expected number of
events with $2$~fb$^{-1}$ at Tevatron drops below 100 (third column), assuming $B(A_H \rightarrow
W^+W^-) = 66\%$. Here $l$ denotes any of $e, \mu, \tau$. }
\label{tab:mqh}
\end{center}\vspace*{-\baselineskip}
\end{table}
%

There are at least three channels where we expect a significant signal for
$m_{q_H} < 300$~GeV, even after experimental cuts have been applied. The
current data from Tevatron therefore strongly suggests that $m_{q_H} >
300$~GeV. It should be possible to derive more stringent bounds with a detailed
analysis and refinement of some of the final states listed above. For instance,
a fraction of the $6j + 2 \, (l \nu_l)$ final
state will have both leptons with the same-sign charge. 

\

Finally we can discuss the case where $q_H$ is only slightly heavier than $A_H$. This is possible only for somewhat larger
values of $f$ where we need to consider $A_H$ decaying into SM gauge bosons. The quarks from
the decay $q_H \rightarrow A_H+q$ are
too soft to be observable in this case. Apart from this, the phenomenology is the same as in the
previous analysis, in particular the
results from Tab.~\ref{tab:mqh} can be adopted by reducing the number of jets by two for each final
state. 
With the same arguments as above we conclude that current experimental data
strongly  suggest $m_{q_H}> 300$~GeV also in this case.

\subsection{LHC phenomenology}
Single production of T-odd particles is possible in principle at LHC, however
the T-violating partial widths are too small for these processes to be
observable. Therefore pair production remains the dominant source of T-odd
particles.  For LHC, the pair production rates,  including the effects of the
mirror fermions, have been  studied in Refs. \cite{Freitas:2006vy} and
\cite{Belyaev:2006jh}. It turns out that not only the T-odd quark production
but also the pair production rates for T-odd gauge bosons depend on the mass
$m_{q_H}$ of the mirror quarks. 

For definiteness, we will take $\kappa = 0.5$ throughout this chapter and comment on other choices later. In this
case the mirror quarks are always somewhat heavier than $W_H$ and $Z_H$. As
above we will consider two cases, case 1 with $f = 750 \gev$ and case 2 with $
f = 1500 \gev$. The corresponding particle masses are:

Case 1: $M_{W_H^\pm} = M_{Z_H} = 482 \gev$, $M_{A_H} = 111 \gev$, $m_{u_H} = 523 \gev$ and $m_{d_H} = 530 \gev$ for the
first and second generation mirror quarks.

Case 2: $M_{W_H^\pm} = M_{Z_H} = 975 \gev$, $M_{A_H} = 235 \gev$, $m_{u_H} = 1057 \gev$ and $m_{d_H} = 1061 \gev$ for the
first and second generation mirror quarks.

\vspace{1em}
\begin{figure}
\psfig{figure=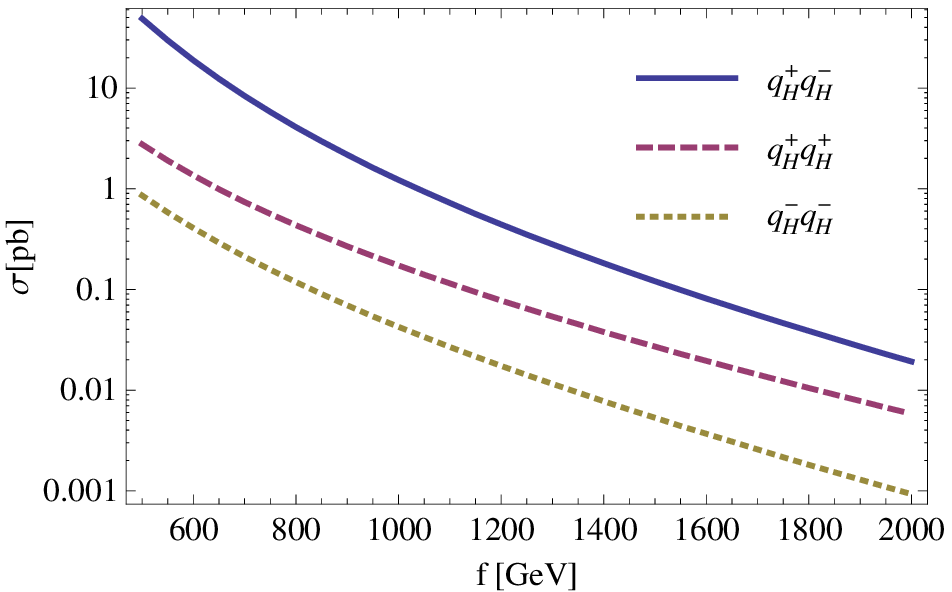, width=0.48\textwidth}\hfill%
{\psfig{figure=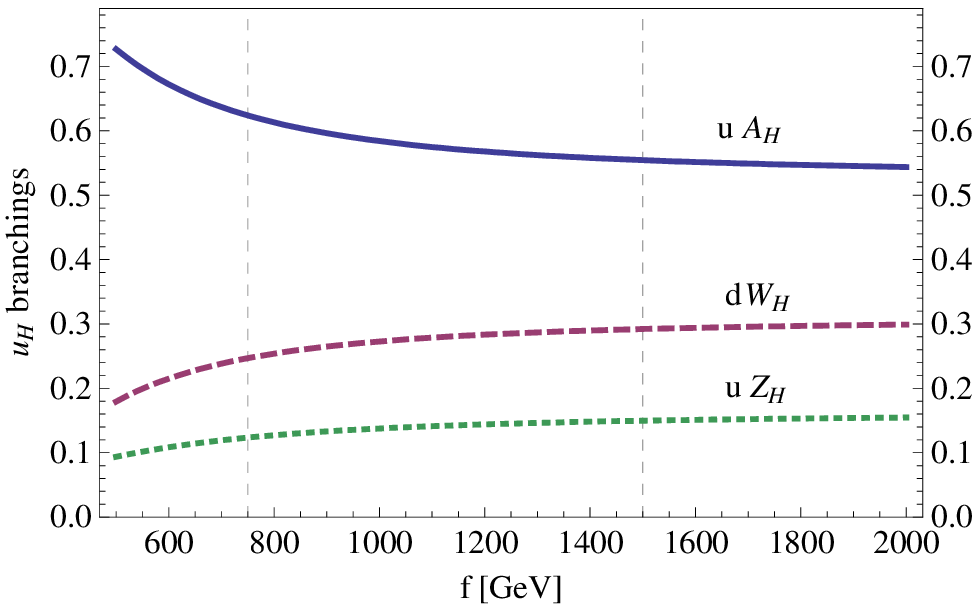, width=0.48\textwidth}}
\mycaption{Left: Production of T-odd quark pairs, where $q^+ =
\{u_H,\bar{d}_H,c_H,\bar{s}_H\}$ and $q^-=\{\bar{u}_H,d_H,\bar{c}_H,s_H\}$, as a
function of $f$ for $\kappa = 0.5$  Right: Branching fractions for the decay of
 $u_H$ with $\kappa = 0.5$. The dashed vertical lines indicate our
two reference scenarios with $f=750$~GeV and $f=1500$~GeV. {The
branching fractions for down-type mirror quarks $d_H$ are similar, although
there are small differences due to ${\cal O}(v^2/f^2)$ mass
corrections~\cite{hubpaz,Blanke:2006eb}.}}
\label{fig:qhlhc}
\end{figure} 
For the chosen value of $\kappa$ the most important source of T-odd particles at
LHC is the pair production of mirror quarks $q_H$. The left plot in Fig.~\ref{fig:qhlhc} 
shows the cross sections for the production of equally charged
and opposite charged mirror quark pairs, for the first two quark families. For
moderate values of $f$ the cross section for $q_H^+q_H^-$ is of the order of one
picobarn. The cross section for positively charged mirror quark pairs is larger
than the one for quark pairs of negative charge and approaches the $q_H^+q_H^-$
cross section for increasing values of $f$. As for the Tevatron calculation, the
renormalization and factorization scales $\mu$  were chosen to be the invariant
mass of the incoming partons. The scale uncertainty is again around 30\%. 

The right plot in Fig.~\ref{fig:qhlhc} shows the branching fractions for the
two-body decays of an up-type mirror quark $u_H$ as a function of $f$. While the decay
into $A_H$ and a SM quark dominates, the branching ratios for the other
channels are sizeable, leading to a large variety of phenomenological
signatures. Note that the mass of the mirror quarks always lies above the
Tevatron bounds of section \ref{ssc:teva} for the considered range of
parameters. 
\begin{table}[t]
\renewcommand{\arraystretch}{1.5}
\begin{center}
    \begin{tabular}{|c||c|c||c|c|}
        \hline 
        \multicolumn{5}{|c|}{\large Signal rates for $f=750 \gev$} \\ \hline\hline 
        $q_H^+q_H^- \rightarrow$ & \multicolumn{2}{c||}{$q^+q^- A_HA_H $ (BF: 39\%)}&
\multicolumn{2}{c|}{$q^+q^+ A_H W_H^- $ (BF: 15\%)} \\\hline 
        & \hspace*{3mm}Final State  \hspace*{3mm}&  $\sigma[\mathrm{fb}]$ & \hspace*{3mm} Final
State\hspace*{3mm} &  $\sigma[\mathrm{fb}]$  \\ \hline 
        & $6j$ & 994        & $6j + l_*^- + \Eslash $ & 124 \\ \hline 
        & $ 4j + \Eslash$ & 568 & $ 4j + l_*^- + \Eslash$ & 71  \\ \hline 
        & $ 4j + ll$ & 319 &$ 4j + l_*^- + ll + \Eslash$ &40 \\ \hline 
        & $ 2j + 2\, ll $ & 26      & $ 2j + l_*^- +  2\, ll +\Eslash$ & 3.2 \\ \hline 
        \hline 
        $q_H^+q_H^- \rightarrow$ & \multicolumn{2}{c||}{$q^-q^+ W_H^-W_H^+ $ (BF: 6\%)}&
\multicolumn{2}{c|}{$q^+q^- Z_H A_H $ (BF: 15\%)} \\\hline 
        &Final State &  $\sigma[\mathrm{fb}]$ & Final State &  $\sigma[\mathrm{fb}]$  \\ \hline 
        & $6j + l_*^+l_*^- + \Eslash $ & 16.0 & $6j + h$ & 306 \\ \hline 
        & $ 4j + ll +  l_*^+l_*^- + \Eslash $ & 5.2 & $4j + ll + h $& 98 \\ \hline 
        & & & $ 4j +h + \Eslash$ & 175 \\ \hline 
\end{tabular}
\mycaption{Signal rates without cuts, from $q^+_H q_H^-$ pair decays. Leptons $l = \{e,\mu,\tau\}$,
and $ll$ always denotes a  charged lepton pair $l^+l^-$ of the same flavor, while a hard charged
lepton coming from  a decay $W_H \rightarrow W A_H  \rightarrow l \nu_l A_H$ is denoted by $l_*$. }
\label{tab:lhc1}
\end{center}
\end{table}
\begin{table}[tp]
\renewcommand{\arraystretch}{1.5}
\begin{center}
    \begin{tabular}{|c||c|c||c|c|}
        \hline 
        \multicolumn{5}{|c|}{\large Signal rates for $f=1500 \gev$} \\ \hline\hline 
        $q_H^+q_H^- \rightarrow$ & \multicolumn{2}{c||}{$q^+q^- A_HA_H $ (BF: 31\%)}&
\multicolumn{2}{c|}{$q^+q^+ A_H W_H^- $ (BF: 16\%)} \\\hline 
        &\hspace*{3mm}Final State\hspace*{3mm} &  $\sigma[\mathrm{fb}]$ & \hspace*{3mm} Final State
\hspace*{3mm} &  $\sigma[\mathrm{fb}]$  \\ \hline 
        & $10 j$ & 8.2 & $10 j + l_*^- + \Eslash$ & 1.37  \\ \hline
        & $ 8j + l + \Eslash$ & 8.4 & $ 8j +l_*^-+ l + \Eslash$ & 1.40 \\ \hline 
        & $ 6j + ll + \Eslash$ & 5.2 & $ 8j + l_*^- + l^- + \Eslash$ & 0.70 \\ \hline 
        & $ 6j + l^\pm l^\pm+ \Eslash$ & 1.6 & $l_*^- + ll + \Eslash + \mathrm{jets}$ &
1.14 \\ \hline 
        \hline 
        $q_H^+q_H^- \rightarrow$ & \multicolumn{2}{c||}{$q^-q^+ W_H^-W_H^+ $ (BF: 9\%)}&
\multicolumn{2}{c|}{$q^+q^- Z_H A_H $ (BF: 17\%)} \\\hline 
        &Final State &  $\sigma[\mathrm{fb}]$ & Final State &  $\sigma[\mathrm{fb}]$  \\ \hline 
        & $10 j +l_*^+ l_*^- + \Eslash$ & 0.25 & $10j + h$ & 3.16 \\ \hline 
        & $ l_*^+l_*^- + l + l + \Eslash  +\mathrm{jets}$ & 0.21 & $6j + h + l^\pm l^\pm +
\Eslash$ & 1.15 \\ \hline 
    \end{tabular}
\end{center}
\mycaption{Signal rates without cuts, from $q^+_H q_H^-$ pair decays. Notation as in
Tab.~\ref{tab:lhc1}.}
\label{tab:lhc2}
\end{table}
We will first discuss the signals stemming from the decay of opposite charge
mirror quark pairs, $q_H^+q_H^-$. When both $q_H$ decay into $ A_H+q$, the
final states are the same as those discussed for Tevatron. The results for case
1 are shown in the upper left block of Tab.~\ref{tab:lhc1}. The cross sections
are large, in particular for the six jet channel, but also the channels with
two or four charged leptons in the final states are well populated. For case 2
the cross sections, shown in Tab.~\ref{tab:lhc2}, are significantly lower. A
detailed analysis would be needed to extract a signal from the background in
this case. 
However we expect very low SM background for the four-lepton channel,
suggesting this signature as a promising discovery channel.

Consider now the case where one of the mirror quarks decays into $W_H+q$, and
the other one into $A_H+q$.  For the values of $f$ considered here the
branching fraction $B( W_H \rightarrow W A_H)$ is above 90\%
\cite{Barger:2007df}, so we will here neglect other decays. We then get
\begin{align}
        q_H^+ q_H^- \longrightarrow q^+ q^+ W^- A_HA_H
\end{align}
as intermediate decay product. We will focus on channels where the $W^-$ decays
leptonically, and denote the corresponding lepton by $l_*^-$. The results for
both cases can be found in the upper right blocks of Tab.~\ref{tab:lhc1} and
Tab.~\ref{tab:lhc2}, respectively.  For case 1, the interesting channels are the same as
before, just with the $l_*^-$ added.  Since the $W^-$ from the $W_H$ decay is
strongly boosted,  a rather strong cut can be imposed on the transverse energy
of $l_*^-$ as well as on the missing transverse energy. This will effectively
reduce the SM background, making these channels suitable for new physics
searches, despite their somewhat smaller signal cross sections. Also for case 2
the cross sections are somewhat smaller than above. However due to the
additional lepton, the same sign dilepton channel is enhanced, and
furthermore the trilepton channel gets a sizeable cross section. 
In addition to the processes in the upper right block of Tab.~\ref{tab:lhc1} and Tab.~\ref{tab:lhc2} also the corresponding charge conjugate final states appear with the same cross sections.
Combining both channels further increases the discovery reach in these decay
modes. 

Even more distinctive final states appear when both mirror quarks decay into $W_H+q$. Here we only consider channels where both $W$ bosons originating from
$W_H$'s decay leptonically. Thus every final state contains two oppositely
charged leptons with uncorrelated flavor. For case 1 the cross sections are
rather small compared to those of the channels considered above, so we only
list the two strongest ones, in the lower left block of Tab.~\ref{tab:lhc1}.
Note that the channel with four leptons in the final state only has a slightly
larger cross section than the five lepton channel from the $A_H W_H$ decay
modes. We therefore do not expect these two channels to be particularly
important for the discovery of the model. For case 2 the situation is similar.
The cross sections for the two most interesting channels are shown in
Tab.~\ref{tab:lhc2}. The second channel has four uncorrelated leptons, and
could for example lead to  $e^+e^+\mu^+ \tau^-$ final states. While this
channel is quite distinctive, it would require a higher luminosity than
presently  forseen.

Finally we consider the case where the mirror quark pair decays into two quarks
and $A_H Z_H$. The novel feature here is that $Z_H$ decays into the Higgs boson
and $A_H$ with a branching fraction of $\sim 80$~\% \cite{Barger:2007df}. We
thus have processes of the form
\begin{align}
        q_H^+ q_H^- \longrightarrow q^+ q^- h A_H A_H. 
\end{align}
The actual value of $B(Z_H \rightarrow h A_H)$ depends on the Higgs boson mass, but
lies above $\sim 80\%$ as long as the $Z_H$ is somewhat heavier than the decay
products, i.e. $M_{Z_H} > m_h+M_{A_H}$. The number of Higgs bosons produced in this channel can be sizeable as illustrated in  Tab.~\ref{tab:lhc1} and
Tab.~\ref{tab:lhc2}, respectively. Depending on the Higgs boson mass, the
(very) final states vary and a more detailed analysis would be required if a 
signal would surface. In Tab.~\ref{tab:lhc1} we list the three channels with
the largest cross section. The production rates for $h + \mathrm{jets}$ and for
$h + \mathrm{jets} + \Eslash$ are sizeable, in total around $0.5\,\,
\mathrm{pb}$. This is similar to $tth$ production in the SM, especially for
larger values of the Higgs mass. The channel with a charged lepton pair
produced along with the Higgs boson is particularly interesting. The signal is
comparable to the SM background coming from $Zh$ production, but can be
distinguished by requiring additional jets from the $q_H$ decays and by the
fact that the lepton pair has an invariant mass $M_{A_H}$. 

As expected, the cross sections are much smaller for case 2. We were still able
to identify an interesting channel where the Higgs is produced along with two
equally charged leptons. While the signal is rather small, the same is true for
the SM background. The results for this channel and for $h + 10\,\,
\mathrm{jets}$ production are shown in the lower right block of Tab.~\ref{tab:lhc2}. 

\

Next we will discuss the signals from decays of positively charged mirror quark pairs,  $q_H^+
q_H^+$. Since the production rate for same sign mirror quark pairs is almost one order of
magnitude smaller than the one for opposite sign mirror quarks, we will only consider decay modes that lead to a distinctive final state. These
mainly come from processes where both mirror quarks decay into $W_H^+$ and a quark, leading to 
\begin{align}
        q_H^+ q_H^+ \longrightarrow q^-q^- W^+W^+ A_H A_H.
\end{align}
To be sensitive to the charges we only consider leptonic decays of the $W$ bosons, leading to two
positively charged hard leptons which we will again denote by $l_*^+$. 

The signal rates for both case 1 and case 2 can be found in Tab.~\ref{tab:lhc3}. 
\begin{table}[tp]
\renewcommand{\arraystretch}{1.5}
\begin{center}
    \begin{tabular}{|c|c||c|c|}
        \hline 
        \multicolumn{4}{|c|}{\large Same Sign Multilepton Rates} \\ \hline \hline 
        \multicolumn{2}{|c||}{$ f= 750 \gev $} &\multicolumn{2}{c|}{$f=1500 \gev$}\\ \hline 
        \hspace*{3mm}Final State\hspace*{3mm} &  $\sigma[\mathrm{fb}]$ & \hspace*{3mm} Final State
\hspace*{3mm} &  $\sigma[\mathrm{fb}]$  \\ \hline 
        $6 j + l^+_*l^+_*$ & 1.56 & $ l^+_*l^+_* + \mathrm{anything}$ & 0.256 \\ \hline
        $ 4j +  l^+_*l^+_* + \Eslash$ & 0.89 & $ l^+_*l^+_* +l^+ + \mathrm{anything}$ & 0.115
\\ \hline
        $ 4j +  l^+_*l^+_*  + ll $ & 0.50 & & \\ \hline 
    \end{tabular}
\end{center}
\mycaption{Rates for same sign lepton signals, from $q^+_H q_H^+$ pair decays. Notation as in
Tab.~\ref{tab:lhc1} and ``$\mathrm{anything}$'' stands for additional jets, leptons and/or missing
energy from neutrinos. }
\label{tab:lhc3}
\end{table}
Since we require leptonic decays of at least two $W$ bosons, the signal rates for both cases are
rather small. With suitable cuts on the transverse momentum of the two hard leptons it should still
be possible to efficiently remove the SM background. 

\vspace{1em}
T-odd gauge bosons in general provide a cleaner signature, since less particles are produced in
their decay. However the cross section for T-odd gauge boson pair production is rather small. For
the values of $f$ and $\kappa$ chosen here at least ten times more T-odd gauge bosons are produced
in the decays of mirror quarks. 

Furthermore it  is hard to distinguish directly produced T-odd gauge bosons from
those coming from mirror quark decays. The reason is that while the possible
final states differ in the total number of jets, that number is rather large for
most processes and thus would require a full reconstruction of all events to
be measured. 

\vspace{1em}
A comment on the effects of a variation of the free parameters
is in order.  If $f$ is increased further, all T-odd particle masses are increased and the cross sections for all processes go down, while the final states and their branching fractions remain essentially unchanged. 

Changing $\kappa$, on the contrary, changes the results more significantly because 
it affects the mirror quark mass $m_{q_H}$ while leaving the heavy gauge boson 
masses unchanged . If $\kappa$ goes below $0.45$, the decays into most T-odd 
gauge bosons become inaccessible,
leaving $A_H+q$ as the only two-body final state. In that case direct T-odd gauge
boson pair production becomes important, since $W_H$ and $Z_H$ are not obtained
from $q_H$ decays anymore. The extreme case where the mirror quarks are also
lighter than $A_H$ will be discussed in the next chapter. On the other hand, if
$\kappa$ is increased, the cross section for $q_Hq_H$ pair production decreases
while at the same time most cross sections for T-odd gauge boson pair production
increase, as has been shown in \cite{Freitas:2006vy}. A further effect is that
the branching fractions of $q_H$ change, with the branching fraction
$B(q_H\rightarrow qW_H)$ reaching around 60\%, while $B(q_H \rightarrow q
A_H)$ drops below the 10\% level.

The T-odd gauge boson pair production rates are affected differently by
increasing $\kappa$. While the cross section for $pp \rightarrow W_H^+ Z_H$
increases, the one for $pp \rightarrow W_H^+ A_H$ is reduced
\cite{Freitas:2006vy}. If the mirror quarks are very heavy and only the T-odd
gauge bosons are accessible, the ratio of these cross sections provides a
possibility to measure $m_{q_H}$ indirectly.

\

While inclusion of the WZW term leads to many new signatures of the Littlest
Higgs model at LHC, none of the processes discussed above is actually sensitive
to its integer parameter $N$. The reason is that vertices containing the WZW
term only appear in decays, and all partial width are multiplied by the same
power of $N$. Measuring the total width of $A_H$ could in principle give access
to $N$, however in practice a width of the order of one $\mathrm{eV}$ is not
measurable. The same problem appears if one tries to measure the T-violating
partial widths in the decays of other T-odd particles like $W_H$. T-violating
decays into two gauge bosons can be distinguished from
T-conserving decays by measuring the distribution of the angle between
the outgoing gauge bosons and the polarization axis of $W_H$
\cite{Krohn:2008ye}. In practice however the T-violating partial width is too
small for this analysis to be feasible. 

Single T-odd gauge boson production via gauge boson fusion would give direct
access to $N$. This process is not in the reach of LHC, but might be detectable
at a very luminous linear collider.

\subsection{The Case of a fermionic LTOP}
\label{f_ltop}

The masses of the mirror fermions are free parameters in the LHT, determined by the parameters
$\kappa_i$. Thus in principle some of them could be lighter than the lightest T-odd gauge
boson $A_H$. With T-parity broken these fermions are unstable, so any of them could be the lightest
T-odd particle (LTOP). 

Since the WZW term breaks T-parity directly only in the gauge sector, the simplest decay process is
a three body decay mediated by a
virtual T-odd gauge boson. Loop induced two body decays can be of similar importance. The most
relevant diagrams are similar to those
shown in Fig.~\ref{fig:ahff}, with the $A_H$ replaced by a $Z$ or $W$ boson and one of the external
fermions replaced by the fermionic LTOP $f_H$. Major decay channels therefore are $f_H \rightarrow Z
f$, $f_H \rightarrow W \tilde{f}$ and $f_H \rightarrow A_H^* f$. The last channel will either yield
a three fermion final state $f'\bar{f}' f $ or a $VV f$ final state with two SM gauge bosons,
depending on the mass of $f_H$ and $A_H$ and the kinematics of the decay. 

We will now briefly discuss how this could affect the phenomenology at hadron colliders. If $f_H$ is
a T-odd quark, it will be pair produced directly in sizeable amounts and decay as discussed above.
Furthermore most of the directly produced T-odd gauge bosons will decay into $f_H f$ pairs, since
the direct decay into SM particles via the WZW term is suppressed in general. If the mirror quark
masses are somewhat hierarchical, one could also imagine longer decay chains, with the branching
fractions depending on the flavor structure of the mirror quark sector. 

Even more appealing is the case where $f_H$ is a lepton. While their direct production rate is small
in that case, the pair produced mirror quarks will now decay via long chains:
\begin{align}
        q_H &\longrightarrow A_H q \longrightarrow l_H^\pm l^\mp q \longrightarrow \dots, \\
        q_H & \longrightarrow W_H q \longrightarrow A_H W q \longrightarrow l_H^\pm  f^\mp  W q
\longrightarrow \dots
\end{align}
Studying these novel
collider signatures of the Littlest Higgs model in more detail would certainly be interesting.

\section{Conclusion}
\label{sc:concl}
In this paper we have considered some phenomenological consequences of the
natural T-parity breaking Wess-Zumino-Witten term in the effective Lagrangian of
the classically T-parity invariant Littlest Higgs model. In particular we
have calculated the loop induced decays of the heavy photon $A_H$ into normal
fermion pairs, assuming $A_H$ to be the lightest T-odd particle. These complement
the known tree level decays into normal gauge bosons \cite{Barger:2007df} and
are the dominant modes for breaking scales below 1 TeV. For these values, the
effect is quite distinct and changes the phenomenology of the model substantially,
because $A_H$ appears as a decay product of any other T-odd particle. 
Due to its small prefactor the Wess-Zumino-Witten term only makes a negligible contribution to electroweak precision observables.

The new  decay modes typically give rise to final states with many jets. 
Comparing with published data from other new physics searches at the Tevatron,
this leads to improved bounds on the mass of the heavy quarks $q_H$ for values of the breaking scale
below 1 TeV. If $q_H$ is heavier than $A_H$, the new
decays of the latter induce new decays of $q_H$. In particular, as illustrated
in Figure~\ref{fig:teva2}, six jet events give rise to a quite strict limit on
$m_{q_H}$ of 350 GeV for $f = 750$ GeV. But the present study gives only a rough
picture and a refined analysis would be necessary for more accurate results.

At the LHC, the situation is even better. For $q_H \bar{q}_H$ pairs , their
decays into light quarks (antiquarks) and $A_H$ would lead to the same final
states as discussed above. But because of the higher energy, larger heavy quark masses can be probed,
where decays of the
form $q_H \to q +W_H$ are possible, leading to  very distinctive finale states
which may be distinguished from SM background. Of particular interest in this
respect are final states with four leptons. Even more tantalizing are processes involving a $Z_H$ boson decaying through a Higgs boson.
The production rate for $h + $jets (and missing energy) from these processes  is sizeable, possibly
comparable to the SM top associated Higgs boson production rate, although highly
dependent on $f$ and the masses $m_{q_H}$. For a moderately heavy Higgs ($m_h \sim
200$ GeV), and a mirror mass around $m_{q_H} \sim 350-400$ GeV as
well as $f \sim 500$ GeV, the 
production rate can be of the order of 10 pb or more. This is important since
other Higgs production channels are reduced in the Littlest Higgs model
\cite{higgsprod}.
The production of two 
equally charged heavy particles, $q_H^\pm q_H^\pm$, is possible,  yielding striking same-sign lepton signatures but small rates. 

In this work we have not considered associated production of heavy quarks with heavy gauge bosons and production of the top quark partners, since pair production of first generation heavy quarks has a substantially larger cross section and thus is more suitable for a first new physics discovery. Nevertheless, a detailed analysis of those processes would be interesting for future work since they could reveal additional information about the model structure. 

Because of the small T-violating branching
ratios, singly produced T-odd particles are practically non-observable and thus
only pair production is phenomenologically relevant.

Finally we have also considered the case of a fermionic lightest T-odd particle.
This would lead to still different signatures that will have to be worked out in
detail. 

\bigskip

\vspace{- .3 cm}
\subsection*{Acknowledgments}
PS wants to thank N. Greiner and Z. Kunszt for useful discussions. 
AF is grateful to Argonne National Laboratory, the University of Chicago and the
Universit\"at Z\"urich for hospitality during stages of this work.
This work was supported in part by the Schweizer Nationalfonds. 


\paragraph*{Note added:} Recently W.~Keung, I.~Low and J.~Shu \cite{Keung:2008ve} proposed a method to distinguish WZW-term induced decays of heavy gauge bosons  from decays mediated by other operators. For large enough $f$ this method can be used to analyse the $A_H \rightarrow ZZ $ decay at LHC. 

\appendix

\section{T-odd gauge invariant counterterms}\label{sec:ct}
{
The counterterms (\ref{ct1}) introduced in section \ref{sc:wzw} are not invariant under the global symmetries or gauge transformations because they were constructed to cancel
divergencies in particular diagrams. In general, couplings between heavy gauge bosons and SM fermions cannot be 
written as covariant derivatives in a fermionic kinetic term. Similarly, counterterms involving T-odd gauge bosons as well as the massive standard model gauge bosons 
also need  to be present; again they do not surface in a symmetric way at first.
We note however that the simple counterterms in  (\ref{ct1}) are only required
to cancel divergencies affecting broken gauge symmetries and 
thus no $U(1)_{em}$ breaking counterterms are required. 

As is well know in chiral perturbation theory, one can do better and construct directly
the required symmetric counterterms, \cite{Gasser:1984ux}. Also in the present case 
we can rewrite the counterterms in a form that preserves gauge invariance and furthermore reflect the underlying $SU(5)$ symmetric structure of the models. The point is to insert the non-linear sigma model field $\Sigma$ into the covariant derivative for the fermions, to define objects with well-defined transformations
properties, such as vectorial and axial currents and consider appropriate powers of them. We do not give a systematic exposition here but refer to a future publication. For the purely mesonic WZW term, such a
complete analysis was given in \cite{Donoghue:1988ct}.

As illustration of the procedure, the counterterms for the diagram in Fig.~\ref{fig:ahff} (b) could have contributions from a term}
\begin{align}
       c_{ct}  \bar{f} Q_W^2 ( \Sigma^\dagger D_\mu  \Sigma - \Sigma D_\mu \Sigma ^\dagger )\gamma^\mu P_L  f.
\end{align}
Here $f = (\psi_1,0,\psi_2)^T$ contains the lefthanded fermion doublets $\psi_1$ and $\psi_2$ that yield the lefthanded SM fermions $\psi_{SM}=\frac{1}{\sqrt{2}}(\psi_1-\psi_2)$ as well as their mirror partners $\psi_H = \frac{1}{\sqrt{2}}(\psi_1+\psi_2)$. The $5\times 5$ matrix $Q_W$ is defined as $Q_W = Q^+ +Q^-$, where $Q^+$ and $Q^-$ are the generators of the $W^+$ and $W^-$ bosons that run in the loop, and $c_{ct}$ is an appropriate coefficient for the counterterm. 

This term is gauge invariant. Upon electroweak symmetry breaking it generates counterterms at order $v^2/f^2$ as required to match the results of Tab.~\ref{tab:ct1}. The righthanded fermions are gauge singlets, so their counterterms have to be generated differently. A possible term is
\begin{align}
        c_{ct}  \bar{f}\gamma^\mu P_R  f \,\mathrm{Tr}\left[ Q_W^2 (\Sigma^\dagger D_\mu  \Sigma - \Sigma D_\mu \Sigma ^\dagger)\right].
\end{align}
Counterterms of this form were introduced in \cite{Savage:1992ac} where they are used to regularize the divergencies in the loop induced decay of the neutral pion $\pi^0$ into lepton pairs, a process that is similar to the decay of $A_H$ into lepton pairs as discussed in section \ref{sc:wzw}. 
\section{Feynman rules for the LHT with WZW term}\label{sec:feynrules}
The Feynman rules for the original Littlest Higgs model  are listed in Ref.~\cite{hanetal}. An almost
complete collection (to leading order in $1/f$) of Feynman rules for the Littlest Higgs model with
T-parity, including flavor effects, can be found in Ref.~\cite{Blanke:2006eb}. 

In Tabs.~\ref{tab:3gb}--\ref{tab:2gb2s3} we provide the additional Feynman rules introduced by the WZW term after anomaly cancellation,
to leading order in $(1/f)$. The
small mixing between $A_H$ and $Z_H$ is neglected since it is numerically less important than other
subleading $(1/f)$ corrections. 

We use the conventions of Ref.~\cite{DennerF} with all momenta incoming. The Feynman rules can be translated to CalcHEP conventions by multiplying each vertex with $(-i)$ and changing $p_i^\mu \rightarrow -p_i^\mu$ for all momenta in the vertex. 
We work in a general covariant gauge. An overall factor $\hat{N}=\frac{N}{48\pi^2}$ has been factored out from
all the Feynman rules. Vertices that are zero have been omitted.

\renewcommand{\arraystretch}{0.97}

\begin{table}[bh]
\begin{center}
\begin{tabular}{|ccc|c|} \hline
\multicolumn{3}{|l|}{\textbf{Particles}} & \textbf{Vertices} \\ \hline
$A_H{}_{\mu }$ & $W^+_{\nu }$ & $W^-_{\rho }$  &
        $-\frac{6}{5}\frac{ e^3  v^2 }{ c_w s_w^2  f^2 }\varepsilon_{\mu \nu \rho \sigma}
\big(p_2^\sigma -p_3^\sigma \big)$\\[2mm] 
$A_H{}_{\mu }$ & ${Z}_{\nu }$ & ${Z}_{\rho }$  &
        $-\frac{6}{5}\frac{ e^3  v^2 }{ c_w^3  s_w^2  f^2 }\varepsilon_{\mu \nu \rho \sigma}
\big(p_2^\sigma -p_3^\sigma \big)$\\[2mm]
${A}_{\mu }$ & $W^+_{\nu }$ & $W_H^-{}_{\rho }$  &
        $\frac{  e^3  v^2 }{ s_w^2  f^2 }\varepsilon_{\mu \nu \rho \sigma} \big(p_2^\sigma
+p_3^\sigma -2p_1^\sigma \big)$\\[2mm]
${A}_{\mu }$ & $W^-_{\nu }$ & $W_H^+{}_{\rho }$  &
        $\frac{ e^3  v^2 }{ s_w^2  f^2 }\varepsilon_{\mu \nu \rho \sigma} \big(p_3^\sigma
+p_2^\sigma -2p_1^\sigma \big)$\\[2mm]
$A_H{}_{\mu }$ & $A_H{}_{\nu }$ & $Z_H{}_{\rho }$  &
        $-\frac{4}{5}\frac{ e^3  v^2 }{ c_w^2  s_w f^2 }\varepsilon_{\mu \nu \rho \sigma}
\big(p_1^\sigma -p_2^\sigma \big)$\\[2mm]
$A_H{}_{\mu }$ & $W_H^+{}_{\nu }$ & $W_H^-{}_{\rho }$  &
        $-\frac{4}{5}\frac{ e^3  v^2 }{ c_w s_w^2  f^2 }\varepsilon_{\mu \nu \rho \sigma}
\big(p_2^\sigma -p_3^\sigma \big)$\\[2mm]
$A_H{}_{\mu }$ & $Z_H{}_{\nu }$ & $Z_H{}_{\rho }$  &
        $\frac{4}{5}\frac{ e^3  v^2 }{ c_w s_w^2  f^2 }\varepsilon_{\mu \nu \rho \sigma}
\big(p_3^\sigma -p_2^\sigma \big)$\\[2mm]
$W^+_{\mu }$ & $W^-_{\nu }$ & $Z_H{}_{\rho }$  &
        $-\frac{ e^3  v^2 }{ s_w^3  f^2 }\varepsilon_{\mu \nu \rho \sigma} \big(p_2^\sigma
-p_1^\sigma \big)$\\[2mm]
$W^+_{\mu }$ & $W_H^-{}_{\nu }$ & ${Z}_{\rho }$  &
        $-\frac{ e^3  v^2 }{ c_w s_w^3  f^2 }\varepsilon_{\mu \nu \rho \sigma} \big( (1+
s_w^2)p_1^\sigma -2 s_w^2 p_3^\sigma - c_w^2 p_2^\sigma \big)$\\[2mm]
$W^-_{\mu }$ & $W_H^+{}_{\nu }$ & ${Z}_{\rho }$  &
        $\frac{ e^3  v^2 }{ c_w s_w^3  f^2 }\varepsilon_{\mu \nu \rho \sigma} \big( c_w^2
p_2^\sigma +2 s_w^2 p_3^\sigma - (1+ s_w^2)p_1^\sigma \big)$\\[2mm]
$W_H^+{}_{\mu }$ & $W_H^-{}_{\nu }$ & $Z_H{}_{\rho }$  &
        $\frac{  e^3  v^2 }{ s_w^3  f^2 }\varepsilon_{\mu \nu \rho \sigma} \big(p_1^\sigma
-p_2^\sigma \big)$\\[2mm]
${Z}_{\mu }$ & ${Z}_{\nu }$ & $Z_H{}_{\rho }$  &
        $-2\frac{e^3  v^2 }{ c_w^2  s_w^3  f^2 }\varepsilon_{\mu \nu \rho \sigma} \big(p_2^\sigma
-p_1^\sigma \big)$\\ \hline
\end{tabular}
\end{center}
\vspace{-3ex}
\mycaption{T-parity violating three gauge boson vertices.
The momenta $p_{1,2,3}$ correspond to the particle in the first, second, and third column,
respectively.
}
\label{tab:3gb}
\end{table}

\begin{table}[tp]
\begin{center}
\begin{tabular}{|cccc|c|} \hline
\multicolumn{4}{|l|}{\textbf{Particles}} & \hspace*{20mm}\textbf{Vertices}\hspace*{20mm} \\ \hline
${A}_{\mu }$  &   $A_H{}_{\nu }$ & $W^+_{\rho }$  &   $W^-_{\sigma }$   &  
        $\frac{12}{5}\frac{ e{}^4  v^2 }{ c_w s_w^2  f^2 }\varepsilon_{\mu \nu \rho \sigma} $\\[2mm]
${A}_{\mu }$  &   $A_H{}_{\nu }$  &   $W_H^+{}_{\rho }$  &   $W_H^-{}_{\sigma }$   &  
        $\frac{8}{5}\frac{ e{}^4  v^2 }{ c_w s_w^2  f^2 }\varepsilon_{\mu \nu \rho \sigma} $\\[2mm]
${A}_{\mu }$  &   $W^+_{\nu }$  &   $W^-_{\rho }$  &   $Z_H{}_{\sigma }$   &  
        $-2\frac{ e{}^4  v^2 }{ s_w^3  f^2 }\varepsilon_{\mu \nu \rho \sigma} $\\[2mm]
${A}_{\mu }$  &   $W^+_{\nu }$  &   $W_H^-{}_{\rho }$  &   ${Z}_{\sigma }$   &  
        $2\frac{ e{}^4  v^2 }{ c_w s_w^3  f^2 }\varepsilon_{\mu \nu \rho \sigma} $\\[2mm]
${A}_{\mu }$  &   $W^-_{\nu }$  &   $W_H^+{}_{\rho }$  &   ${Z}_{\sigma }$   &  
        $-2\frac{ e{}^4  v^2 }{ c_w s_w^3  f^2 }\varepsilon_{\mu \nu \rho \sigma} $\\[2mm]
${A}_{\mu }$  &   $W_H^+{}_{\nu }$  &   $W_H^-{}_{\rho }$  &   $Z_H{}_{\sigma }$   &  
        $-2\frac{ e{}^4  v^2 }{ s_w^3  f^2 }\varepsilon_{\mu \nu \rho \sigma} $\\[2mm]
$A_H{}_{\mu }$  &   $W^+_{\nu }$  &   $W^-_{\rho }$  &   ${Z}_{\sigma }$   &  
        $-\frac{6}{5}\frac{ (3-2 s_w ^2) e{}^4  v^2 }{ c_w^2  s_w^3  f^2 }\varepsilon_{\mu \nu \rho \sigma} $\\[2mm]
$A_H{}_{\mu }$  &   $W^+_{\nu }$  &   $W_H^-{}_{\rho }$  &   $Z_H{}_{\sigma }$   &  
        $-\frac{4}{5}\frac{ e{}^4  v^2 }{ c_w s_w^3  f^2 }\varepsilon_{\mu \nu \rho \sigma} $\\[2mm]
$A_H{}_{\mu }$  &   $W^-_{\nu }$  &   $W_H^+{}_{\rho }$  &   $Z_H{}_{\sigma }$   &  
        $\frac{4}{5}\frac{ e{}^4  v^2 }{ c_w s_w^3  f^2 }\varepsilon_{\mu \nu \rho \sigma} $\\[2mm]
$A_H{}_{\mu }$  &   $W_H^+{}_{\nu }$  &   $W_H^-{}_{\rho }$  &   ${Z}_{\sigma }$   &  
        $-\frac{4}{5}\frac{ (1-2 s_w ^2) e{}^4  v^2 }{ c_w^2  s_w^3  f^2 }\varepsilon_{\mu \nu \rho \sigma} $\\[2mm]
$W^+_{\mu }$  &   $W^-_{\nu }$  &   ${Z}_{\rho }$  &   $Z_H{}_{\sigma }$   &  
        $-2\frac{ (2- s_w ^2) e{}^4  v^2 }{ c_w s_w^4  f^2 }\varepsilon_{\mu \nu \rho \sigma} $\\[2mm]
$W_H^+{}_{\mu }$  &   $W_H^-{}_{\nu }$  &   ${Z}_{\rho }$  &   $Z_H{}_{\sigma }$   &  
        $-2\frac{ c_w e{}^4  v^2 }{ s_w^4  f^2 }\varepsilon_{\mu \nu \rho \sigma} $\\[2pt] \hline
\end{tabular}
\end{center}
\vspace{-3ex}
\mycaption{T-parity violating vertices with four gauge bosons. 
}
\label{tab:4gb}
\end{table}

\begin{table}[tp]
\begin{center}
\begin{tabular}{|ccc|c|} \hline
\multicolumn{3}{|l|}{\textbf{Particles}} & \textbf{Vertices} \\ \hline
$W^+_{\mu }$  &  $W_H^-{}_{\nu }$  &  ${h}_{}$  & 
        $\frac{ e^2  v}{ s_w^2  f^2 } \varepsilon_{\mu \nu  \rho\sigma} \big(p_2^\rho -p_1^\rho \big)p_3^\sigma$\\[2mm]
$W^-_{\mu }$  &  $W_H^+{}_{\nu }$  &  ${h}_{}$  & 
        $\frac{ e^2  v}{ s_w^2  f^2 } \varepsilon_{\mu \nu  \rho\sigma} \big(p_1^\rho -p_2^\rho \big)p_3^\sigma$ \\[2mm] \hline
\end{tabular}
\end{center}
\vspace{-3ex}
\mycaption{T-parity violating vertices with one physical scalar and two gauge bosons. 
The momenta $p_{1,2,3}$ correspond to the particle in the first, second, and third column,
respectively.
}
\label{tab:2gb1s}
\end{table}

\begin{table}[tp]
\begin{center}
\begin{tabular}{|cccc|c|} \hline
\multicolumn{4}{|l|}{\textbf{Particles}} & \textbf{Vertices} \\ \hline
${A}_{\mu }$&$W^+_{\nu }$&$W_H^-{}_{\rho }$&${h}_{}$ & 
        $2-\frac{ e^3  v}{ s_w^2  f^2 }\varepsilon_{\mu \nu \rho \sigma} \big(p_2 ^\sigma +p_3 ^\sigma -2p_1 ^\sigma \big)$\\[2mm]
${A}_{\mu }$&$W^-_{\nu }$&$W_H^+{}_{\rho }$&${h}_{}$ & 
        $-2\frac{ e^3  v}{ s_w^2  f^2 }\varepsilon_{\mu \nu \rho \sigma} \big(p_3 ^\sigma +p_2 ^\sigma -2p_1 ^\sigma \big)$\\[2mm]
$A_H{}_{\mu }$&$A_H{}_{\nu }$&$Z_H{}_{\rho }$&${h}_{}$ & 
        $\frac{8}{5}\frac{ e^3  v}{ c_w^2  s_w f^2 }\varepsilon_{\mu \nu \rho \sigma} \big(p_1 ^\sigma -p_2 ^\sigma \big)$\\[2mm]
$A_H{}_{\mu }$&$W^+_{\nu }$&$W^-_{\rho }$&${h}_{}$ & 
        $\frac{12}{5}\frac{ e^3  v}{ c_w s_w^2  f^2 }\varepsilon_{\mu \nu \rho \sigma} \big(p_2 ^\sigma -p_3 ^\sigma \big)$\\[2mm]
$A_H{}_{\mu }$&$W_H^+{}_{\nu }$&$W_H^-{}_{\rho }$&${h}_{}$ & 
        $\frac{8}{5}\frac{ e^3  v}{ c_w s_w^2  f^2 }\varepsilon_{\mu \nu \rho \sigma} \big(p_2 ^\sigma -p_3 ^\sigma \big)$\\[2mm]
$A_H{}_{\mu }$&${Z}_{\nu }$&${Z}_{\rho }$&${h}_{}$ & 
        $-\frac{12}{5}\frac{ e^3  v}{ c_w^3  s_w^2  f^2 }\varepsilon_{\mu \nu \rho \sigma} \big(p_3 ^\sigma -p_2 ^\sigma \big)$\\[2mm]
$A_H{}_{\mu }$&$Z_H{}_{\nu }$&$Z_H{}_{\rho }$&${h}_{}$ & 
        $\frac{8}{5}\frac{ e^3  v}{ c_w s_w^2  f^2 }\varepsilon_{\mu \nu \rho \sigma} \big(p_2 ^\sigma -p_3 ^\sigma \big)$\\[2mm]
$W^+_{\mu }$&$W^-_{\nu }$&$Z_H{}_{\rho }$&${h}_{}$ & 
        $2\frac{ e^3  v}{ s_w^3  f^2 }\varepsilon_{\mu \nu \rho \sigma} \big(p_2 ^\sigma -p_1 ^\sigma \big)$\\[2mm]
$W^+_{\mu }$&$W_H^-{}_{\nu }$&${Z}_{\rho }$&${h}_{}$ & 
        $2\frac{ e^3  v}{ c_w s_w^3  f^2 }\varepsilon_{\mu \nu \rho \sigma} \big( (1+ s_w ^2)p_1 ^\sigma -2 s_w^2 p_3 ^\sigma - c_w^2 p_2 ^\sigma \big)$\\[2mm]
$W^-_{\mu }$&$W_H^+{}_{\nu }$&${Z}_{\rho }$&${h}_{}$ & 
        $-2\frac{ e^3  v}{ c_w s_w^3  f^2 }\varepsilon_{\mu \nu \rho \sigma} \big( c_w^2 p_2 ^\sigma +2 s_w^2 p_3 ^\sigma - (1+ s_w ^2)p_1 ^\sigma \big)$\\[2mm]
$W_H^+{}_{\mu }$&$W_H^-{}_{\nu }$&$Z_H{}_{\rho }$&${h}_{}$ & 
        $-2\frac{ e^3  v}{ s_w^3  f^2 }\varepsilon_{\mu \nu \rho \sigma} \big(p_1 ^\sigma -p_2 ^\sigma \big)$\\[2mm]
${Z}_{\mu }$&${Z}_{\nu }$&$Z_H{}_{\rho }$&${h}_{}$ & 
        $4\frac{ e^3  v}{ c_w^2  s_w^3  f^2 }\varepsilon_{\mu \nu \rho \sigma} \big(p_2 ^\sigma -p_1 ^\sigma \big)$\\[2mm] \hline
\end{tabular}
\end{center}
\vspace{-3ex}
\mycaption{T-parity violating vertices with one physical scalar and three gauge bosons. 
The momenta $p_{1,2,3,4}$ correspond to the particle in the first, second, third and fourth column,
respectively.
}
\label{tab:3gb1s}
\end{table}

\begin{table}[tp]
\begin{center}
\begin{tabular}{|c@{\hspace{0.7em}}c@{\hspace{0.7em}}c|c|} \hline
\multicolumn{3}{|l|}{\textbf{Particles}} & \textbf{Vertices} \\ \hline
$A_H{}_{\mu }$ & $W^\pm_{\nu }$ & $G^\mp$   &
        $\mp\frac{6}{5}\frac{ e^2 v }{ c_w s_w  f^2 }\varepsilon_{\mu \nu \rho \sigma} \, p_3^\rho
\big(p_1^\sigma -p_2^\sigma \big)$\\[2mm] 
$A_H{}_{\mu }$ & ${Z}_{\nu }$ & $G^0$   &
        $-\frac{6}{5}\frac{ i e^2 v }{ c_w^2  s_w  f^2 }\varepsilon_{\mu \nu \rho \sigma} \, p_3^\rho
\big(p_1^\sigma -p_2^\sigma \big)$\\[2mm]
$W_H^\pm{}_{\mu }$ & $A_{\nu }$ & $G^\mp$   &
        $\mp 3\frac{ e^2 v }{ s_w  f^2 }\varepsilon_{\mu \nu \rho \sigma} \, p_3^\rho
\big(p_1^\sigma -p_2^\sigma \big)$\\[2mm]
$Z_H{}_{\mu }$ & $W^\pm_{\nu }$ & $G^\mp$   &
        $\pm\frac{ e^2 v }{ s_w^2  f^2 }\varepsilon_{\mu \nu \rho \sigma} \, p_3^\rho
\big(p_1^\sigma -p_2^\sigma \big)$\\[2mm]
$W_H^\pm{}_{\mu }$ & $Z_{\nu }$ & $G^\mp$   &
        $\mp\frac{ e^2 (c_w^2-2 s_w^2) v }{ c_w s_w^2  f^2 }\varepsilon_{\mu \nu \rho \sigma} \,
p_3^\rho \big(p_1^\sigma -p_2^\sigma \big)$\\[2mm]
$W_H^\pm{}_{\mu }$ & $W^\pm_{\nu }$ & $G^0$   &
        $2\frac{ i e^2 v }{ s_w^2  f^2 }\varepsilon_{\mu \nu \rho \sigma} \, p_3^\rho \big(p_1^\sigma
-p_2^\sigma \big)$\\[2mm]
$Z_H{}_{\mu }$ & ${Z}_{\nu }$ & $G^0$   &
        $2\frac{ i e^2 v }{ c_w  s_w^2  f^2 }\varepsilon_{\mu \nu \rho \sigma} \, p_3^\rho
\big(p_1^\sigma -p_2^\sigma \big)$\\ \hline
\end{tabular}
\end{center}
\vspace{-3ex}
\mycaption{T-parity violating vertices with one SM Goldstone  and two gauge bosons.
The momenta $p_{1,2,3}$ correspond to the particle in the first, second, and third column,
respectively.%
}
\label{tab:gold}
\end{table}

\begin{table}[tp]
\begin{center}
\begin{tabular}{|cccc|c|} \hline
\multicolumn{4}{|l|}{\textbf{Particles}} & \textbf{Vertices} \\ \hline
${A}_{\mu }$  &  $A_H{}_{\nu }$  &  $\Phi^+{}_{}$  &  $\Phi^-{}_{}$   & 
        $\frac{2}{5}\frac{ e^2 }{ c_w f^2 }\varepsilon_{\mu \nu \rho \sigma} \big(p_1^\rho p_4^\sigma -p_2^\rho p_4^\sigma -p_1^\rho p_3^\sigma +p_2^\rho p_3^\sigma -2p_3^\rho p_4^\sigma \big)$\\[2mm]
${A}_{\mu }$  &  $A_H{}_{\nu }$  &  $\Phi^{++}{}_{}$  &  $\Phi^{--}{}_{}$   & 
        $\frac{4}{5}\frac{ e^2 }{ c_w f^2 }\varepsilon_{\mu \nu \rho \sigma} \big(p_1^\rho p_4^\sigma -p_2^\rho p_4^\sigma -p_1^\rho p_3^\sigma +p_2^\rho p_3^\sigma -2p_3^\rho p_4^\sigma \big)$\\[2mm]
${A}_{\mu }$  &  $W_H^+{}_{\nu }$  &  $\Phi^+{}_{}$  &  $\Phi^{--}{}_{}$   & 
        $-2\frac{ e^2 }{ s_w f^2 }\varepsilon_{\mu \nu \rho \sigma} \big(p_2^\rho p_4^\sigma -p_1^\rho p_3^\sigma +p_2^\rho p_3^\sigma -p_1^\rho p_4^\sigma \big)$\\[2mm]
${A}_{\mu }$  &  $W_H^+{}_{\nu }$  &  $\Phi^-{}_{}$  &  $\Phi^0{}_{}$   & 
        $\frac{ e^2  \cdot\sqrt{2}}{ s_w f^2 }\varepsilon_{\mu \nu \rho \sigma} \big(3p_1^\rho p_3^\sigma +p_2^\rho p_4^\sigma -p_1^\rho p_4^\sigma -3p_2^\rho p_3^\sigma +4p_3^\rho p_4^\sigma \big)$\\[2mm]
${A}_{\mu }$  &  $W_H^+{}_{\nu }$  &  $\Phi^-{}_{}$  &  $\Phi^p{}_{}$   & 
        $\frac{ i e^2  \cdot\sqrt{2}}{ s_w f^2 }\varepsilon_{\mu \nu \rho \sigma} \big(3p_1^\rho p_3^\sigma +p_2^\rho p_4^\sigma -p_1^\rho p_4^\sigma -3p_2^\rho p_3^\sigma +4p_3^\rho p_4^\sigma \big)$\\[2mm]
${A}_{\mu }$  &  $W_H^-{}_{\nu }$  &  $\Phi^+{}_{}$  &  $\Phi^0{}_{}$   & 
        $-\frac{ e^2  \cdot\sqrt{2}}{ s_w f^2 }\varepsilon_{\mu \nu \rho \sigma} \big(p_2^\rho p_4^\sigma +3p_1^\rho p_3^\sigma -3p_2^\rho p_3^\sigma -p_1^\rho p_4^\sigma +4p_3^\rho p_4^\sigma \big)$\\[2mm]
${A}_{\mu }$  &  $W_H^-{}_{\nu }$  &  $\Phi^+{}_{}$  &  $\Phi^p{}_{}$   & 
        $\frac{ i e^2  \cdot\sqrt{2}}{ s_w f^2 }\varepsilon_{\mu \nu \rho \sigma} \big(p_2^\rho p_4^\sigma +3p_1^\rho p_3^\sigma -3p_2^\rho p_3^\sigma -p_1^\rho p_4^\sigma +4p_3^\rho p_4^\sigma \big)$\\[2mm]
${A}_{\mu }$  &  $W_H^-{}_{\nu }$  &  $\Phi^{++}{}_{}$  &  $\Phi^-{}_{}$   & 
        $-2\frac{ e^2 }{ s_w f^2 }\varepsilon_{\mu \nu \rho \sigma} \big(p_1^\rho p_4^\sigma -p_2^\rho p_3^\sigma +p_1^\rho p_3^\sigma -p_2^\rho p_4^\sigma \big)$\\[2mm]
${A}_{\mu }$  &  $Z_H{}_{\nu }$  &  $\Phi^+{}_{}$  &  $\Phi^-{}_{}$   & 
        $6\frac{ e^2 }{ s_w f^2 }\varepsilon_{\mu \nu \rho \sigma} \big(p_1^\rho p_4^\sigma -p_2^\rho p_4^\sigma -p_1^\rho p_3^\sigma +p_2^\rho p_3^\sigma -2p_3^\rho p_4^\sigma \big)$\\[2mm]
${A}_{\mu }$  &  $Z_H{}_{\nu }$  &  $\Phi^{++}{}_{}$  &  $\Phi^{--}{}_{}$   & 
        $4\frac{ e^2 }{ s_w f^2 }\varepsilon_{\mu \nu \rho \sigma} \big(p_1^\rho p_4^\sigma -p_2^\rho p_4^\sigma -p_1^\rho p_3^\sigma +p_2^\rho p_3^\sigma -2p_3^\rho p_4^\sigma \big)$\\[2mm]
$A_H{}_{\mu }$  &  $W^+_{\nu }$  &  $\Phi^+{}_{}$  &  $\Phi^{--}{}_{}$   & 
        $\frac{2}{5}\frac{ e^2 }{ c_w s_w f^2 }\varepsilon_{\mu \nu \rho \sigma} \big(p_1^\rho p_4^\sigma -p_2^\rho p_4^\sigma +p_2^\rho p_3^\sigma -p_1^\rho p_3^\sigma +2p_3^\rho p_4^\sigma \big)$\\[2mm]
$A_H{}_{\mu }$  &  $W^+_{\nu }$  &  $\Phi^-{}_{}$  &  $\Phi^0{}_{}$   & 
        $\frac{1}{5}\frac{ e^2  \cdot\sqrt{2}}{ c_w s_w f^2 }\varepsilon_{\mu \nu \rho \sigma} \big(p_1^\rho p_3^\sigma -p_2^\rho p_3^\sigma +p_2^\rho p_4^\sigma -p_1^\rho p_4^\sigma -2p_3^\rho p_4^\sigma \big)$\\[2mm]
$A_H{}_{\mu }$  &  $W^+_{\nu }$  &  $\Phi^-{}_{}$  &  $\Phi^p{}_{}$   & 
        $\frac{1}{5}\frac{ i e^2  \cdot\sqrt{2}}{ c_w s_w f^2 }\varepsilon_{\mu \nu \rho \sigma} \big(p_1^\rho p_3^\sigma -p_2^\rho p_3^\sigma +p_2^\rho p_4^\sigma -p_1^\rho p_4^\sigma -2p_3^\rho p_4^\sigma \big)$\\[2mm]
$A_H{}_{\mu }$  &  $W^-_{\nu }$  &  $\Phi^+{}_{}$  &  $\Phi^0{}_{}$   & 
        $-\frac{1}{5}\frac{ e^2  \cdot\sqrt{2}}{ c_w s_w f^2 }\varepsilon_{\mu \nu \rho \sigma} \big(p_2^\rho p_4^\sigma -p_1^\rho p_4^\sigma +p_1^\rho p_3^\sigma -p_2^\rho p_3^\sigma -2p_3^\rho p_4^\sigma \big)$\\[2mm]
$A_H{}_{\mu }$  &  $W^-_{\nu }$  &  $\Phi^+{}_{}$  &  $\Phi^p{}_{}$   & 
        $\frac{1}{5}\frac{ i e^2  \cdot\sqrt{2}}{ c_w s_w f^2 }\varepsilon_{\mu \nu \rho \sigma} \big(p_2^\rho p_4^\sigma -p_1^\rho p_4^\sigma +p_1^\rho p_3^\sigma -p_2^\rho p_3^\sigma -2p_3^\rho p_4^\sigma \big)$\\[2mm]
$A_H{}_{\mu }$  &  $W^-_{\nu }$  &  $\Phi^{++}{}_{}$  &  $\Phi^-{}_{}$   & 
        $-\frac{2}{5}\frac{ e^2 }{ c_w s_w f^2 }\varepsilon_{\mu \nu \rho \sigma} \big(p_2^\rho p_4^\sigma -p_1^\rho p_4^\sigma +p_1^\rho p_3^\sigma -p_2^\rho p_3^\sigma -2p_3^\rho p_4^\sigma \big)$\\[2mm]
$A_H{}_{\mu }$  &  ${Z}_{\nu }$  &  $\Phi^+{}_{}$  &  $\Phi^-{}_{}$   & 
        $-\frac{2}{5}\frac{ e^2  s_w}{ c_w^2  f^2 }\varepsilon_{\mu \nu \rho \sigma} \big(p_1^\rho p_4^\sigma -p_2^\rho p_4^\sigma -p_1^\rho p_3^\sigma +p_2^\rho p_3^\sigma +2p_3^\rho p_4^\sigma \big)$\\[2mm]
$A_H{}_{\mu }$  &  ${Z}_{\nu }$  &  $\Phi^{++}{}_{}$  &  $\Phi^{--}{}_{}$   & 
        $\frac{2}{5}\frac{ (1-2 s_w ^2) e^2 }{ c_w^2  s_w f^2 }\varepsilon_{\mu \nu \rho \sigma} \big(p_1^\rho p_4^\sigma -p_2^\rho p_4^\sigma -p_1^\rho p_3^\sigma +p_2^\rho p_3^\sigma +2p_3^\rho p_4^\sigma \big)$\\[2mm]
$A_H{}_{\mu }$  &  ${Z}_{\nu }$  &  $\Phi^0{}_{}$  &  $\Phi^p{}_{}$   & 
        $-\frac{2}{5}\frac{ i e^2 }{ c_w^2  s_w f^2 }\varepsilon_{\mu \nu \rho \sigma} \big(p_1^\rho p_3^\sigma -p_2^\rho p_3^\sigma -p_1^\rho p_4^\sigma +p_2^\rho p_4^\sigma -2p_3^\rho p_4^\sigma \big)$\\[2mm]
$W^+_{\mu }$  &  $W_H^+{}_{\nu }$  &  $\Phi^{--}{}_{}$  &  $\Phi^0{}_{}$   & 
        $-2\frac{ e^2  \cdot\sqrt{2}}{ s_w^2  f^2 }\varepsilon_{\mu \nu \rho \sigma} \big(p_1^\rho p_3^\sigma -p_2^\rho p_3^\sigma +p_2^\rho p_4^\sigma -p_1^\rho p_4^\sigma +2p_3^\rho p_4^\sigma \big)$\\[2mm]
$W^+_{\mu }$  &  $W_H^+{}_{\nu }$  &  $\Phi^{--}{}_{}$  &  $\Phi^p{}_{}$   & 
        $-2\frac{ i e^2  \cdot\sqrt{2}}{ s_w^2  f^2 }\varepsilon_{\mu \nu \rho \sigma} \big(p_1^\rho p_3^\sigma -p_2^\rho p_3^\sigma +p_2^\rho p_4^\sigma -p_1^\rho p_4^\sigma +2p_3^\rho p_4^\sigma \big)$\\[2mm]
$W^+_{\mu }$  &  $W_H^-{}_{\nu }$  &  ${h}_{}$  &  ${h}_{}$   & 
        $\frac{ e^2 }{ s_w^2  f^2 }\varepsilon_{\mu \nu \rho \sigma} \big(p_2^\rho p_3^\sigma -p_1^\rho p_3^\sigma +p_2^\rho p_4^\sigma -p_1^\rho p_4^\sigma \big)$\\[2mm]
$W^+_{\mu }$  &  $W_H^-{}_{\nu }$  &  $\Phi^p{}_{}$  &  $\Phi^p{}_{}$   & 
        $-2\frac{ e^2 }{ s_w^2  f^2 }\varepsilon_{\mu \nu \rho \sigma} \big(p_1^\rho p_3^\sigma -p_2^\rho p_3^\sigma +p_1^\rho p_4^\sigma -p_2^\rho p_4^\sigma \big)$\\[2mm]\hline 
\end{tabular}
\end{center}
\vspace{-1ex}
\mycaption{T-parity violating vertices with two scalars and two gauge bosons. 
The momenta $p_{1,2,3,4}$ correspond to the particle in the first, second, third and fourth column,
respectively.
{\rm Continued in Tabs.~\ref{tab:2gb2s2}
and \ref{tab:2gb2s3}}}
\label{tab:2gb2s}
\end{table}

\begin{table}[tp]
\begin{center}
\begin{tabular}{|cccc|c|} \hline
\multicolumn{4}{|l|}{\textbf{Particles}} & \textbf{Vertices} \\ \hline
$W^+_{\mu }$  &  $Z_H{}_{\nu }$  &  $\Phi^+{}_{}$  &  $\Phi^{--}{}_{}$   & 
        $4\frac{ e^2 }{ s_w^2  f^2 }\varepsilon_{\mu \nu \rho\sigma} \, p_4^\sigma  \big(p_2^\rho -p_1^\rho +p_3^\rho \big)$\\[2mm]
$W^+_{\mu }$  &  $Z_H{}_{\nu }$  &  $\Phi^-{}_{}$  &  $\Phi^0{}_{}$   & 
        $-2\frac{ e^2  \cdot\sqrt{2}}{ s_w^2  f^2 }\varepsilon_{\mu \nu  \rho\sigma} \, p_4^\sigma  \big(p_1^\rho -p_2^\rho -p_3^\rho \big)$\\[2mm]
$W^+_{\mu }$  &  $Z_H{}_{\nu }$  &  $\Phi^-{}_{}$  &  $\Phi^p{}_{}$   & 
        $-2\frac{ i e^2  \cdot\sqrt{2}}{ s_w^2  f^2 }\varepsilon_{\mu \nu \rho\sigma} \, p_4^\sigma  \big(p_1^\rho -p_2^\rho -p_3^\rho \big)$\\[2mm]
$W^-_{\mu }$  &  $W_H^+{}_{\nu }$  &  ${h}_{}$  &  ${h}_{}$   & 
        $\frac{ e^2 }{ s_w^2  f^2 }\varepsilon_{\mu \nu \rho \sigma} \big(p_1^\rho p_3^\sigma -p_2^\rho p_3^\sigma +p_1^\rho p_4^\sigma -p_2^\rho p_4^\sigma \big)$\\[2mm]
$W^-_{\mu }$  &  $W_H^+{}_{\nu }$  &  $\Phi^+{}_{}$  &  $\Phi^-{}_{}$   & 
        $2\frac{ e^2 }{ s_w^2  f^2 }\varepsilon_{\mu \nu \rho \sigma} \big(p_1^\rho p_4^\sigma -p_2^\rho p_4^\sigma +p_2^\rho p_3^\sigma -p_1^\rho p_3^\sigma -2p_3^\rho p_4^\sigma \big)$\\[2mm]
$W^-_{\mu }$  &  $W_H^+{}_{\nu }$  &  $\Phi^{++}{}_{}$  &  $\Phi^{--}{}_{}$   & 
        $-2\frac{ e^2 }{ s_w^2  f^2 }\varepsilon_{\mu \nu \rho \sigma} \big(p_2^\rho p_4^\sigma +3p_1^\rho p_3^\sigma -3p_2^\rho p_3^\sigma -p_1^\rho p_4^\sigma +4p_3^\rho p_4^\sigma \big)$\\[2mm]
$W^-_{\mu }$  &  $W_H^+{}_{\nu }$  &  $\Phi^0{}_{}$  &  $\Phi^0{}_{}$   & 
        $-2\frac{ e^2 }{ s_w^2  f^2 }\varepsilon_{\mu \nu \rho \sigma} \big(p_2^\rho p_3^\sigma -p_1^\rho p_3^\sigma +p_2^\rho p_4^\sigma -p_1^\rho p_4^\sigma \big)$\\[2mm]
$W^-_{\mu }$  &  $W_H^+{}_{\nu }$  &  $\Phi^0{}_{}$  &  $\Phi^p{}_{}$   & 
        $4\frac{ i e^2 }{ s_w^2  f^2 }\varepsilon_{\mu \nu \rho \sigma} \big(p_1^\rho p_3^\sigma -p_1^\rho p_4^\sigma +p_2^\rho p_4^\sigma -p_2^\rho p_3^\sigma +2p_3^\rho p_4^\sigma \big)$\\[2mm]
$W^-_{\mu }$  &  $W_H^+{}_{\nu }$  &  $\Phi^p{}_{}$  &  $\Phi^p{}_{}$   & 
        $-2\frac{ e^2 }{ s_w^2  f^2 }\varepsilon_{\mu \nu \rho \sigma} \big(p_2^\rho p_3^\sigma -p_1^\rho p_3^\sigma +p_2^\rho p_4^\sigma -p_1^\rho p_4^\sigma \big)$\\[2mm]
$W^-_{\mu }$  &  $W_H^-{}_{\nu }$  &  $\Phi^{++}{}_{}$  &  $\Phi^0{}_{}$   & 
        $-2\frac{ e^2  \cdot\sqrt{2}}{ s_w^2  f^2 }\varepsilon_{\mu \nu \rho \sigma} \big(p_1^\rho p_4^\sigma -p_2^\rho p_4^\sigma +p_2^\rho p_3^\sigma -p_1^\rho p_3^\sigma -2p_3^\rho p_4^\sigma \big)$\\[2mm]
$W^-_{\mu }$  &  $W_H^-{}_{\nu }$  &  $\Phi^{++}{}_{}$  &  $\Phi^p{}_{}$   & 
        $2\frac{ i e^2  \cdot\sqrt{2}}{ s_w^2  f^2 }\varepsilon_{\mu \nu \rho \sigma} \big(p_1^\rho p_4^\sigma -p_2^\rho p_4^\sigma +p_2^\rho p_3^\sigma -p_1^\rho p_3^\sigma -2p_3^\rho p_4^\sigma \big)$\\[2mm]
$W^-_{\mu }$  &  $Z_H{}_{\nu }$  &  $\Phi^+{}_{}$  &  $\Phi^0{}_{}$   & 
        $2\frac{ e^2  \cdot\sqrt{2}}{ s_w^2  f^2 }\varepsilon_{\mu \nu \rho\sigma} \, p_4^\sigma  \big(p_1^\rho -p_2^\rho -p_3^\rho \big)$\\[2mm]
$W^-_{\mu }$  &  $Z_H{}_{\nu }$  &  $\Phi^+{}_{}$  &  $\Phi^p{}_{}$   & 
        $-2\frac{ i e^2  \cdot\sqrt{2}}{ s_w^2  f^2 }\varepsilon_{\mu \nu \rho\sigma} \, p_4^\sigma  \big(p_1^\rho -p_2^\rho -p_3^\rho \big)$\\[2mm]
$W^-_{\mu }$  &  $Z_H{}_{\nu }$  &  $\Phi^{++}{}_{}$  &  $\Phi^-{}_{}$   & 
        $-4\frac{ e^2 }{ s_w^2  f^2 }\varepsilon_{\mu \nu \rho \sigma} \big(p_2^\rho p_3^\sigma -p_1^\rho p_3^\sigma -p_3^\rho p_4^\sigma \big)$\\[2mm]
$W_H^+{}_{\mu }$  &  ${Z}_{\nu }$  &  $\Phi^+{}_{}$  &  $\Phi^{--}{}_{}$   & 
        $2\frac{ e^2 }{ c_w s_w^2  f^2 }\varepsilon_{\mu \nu \rho \sigma} \big( s_w^2 p_2^\rho p_4^\sigma - s_w^2 p_1^\rho p_4^\sigma + (2- s_w ^2)p_1^\rho p_3^\sigma $ \\[2mm]
  &&&& $- (2- s_w ^2)p_2^\rho p_3^\sigma -2p_3^\rho p_4^\sigma \big)$\\[2mm]
$W_H^+{}_{\mu }$  &  ${Z}_{\nu }$  &  $\Phi^-{}_{}$  &  $\Phi^0{}_{}$   & 
        $-\frac{ e^2  \cdot\sqrt{2}}{ c_w s_w^2  f^2 }\varepsilon_{\mu \nu \rho \sigma} \big( (2-3 s_w ^2)p_2^\rho p_3^\sigma - (2-3 s_w ^2)p_1^\rho p_3^\sigma $ \\[2mm]
  &&&& $- s_w^2 p_1^\rho p_4^\sigma + s_w^2 p_2^\rho p_4^\sigma +2 (1-2 s_w ^2)p_3^\rho p_4^\sigma \big)$\\[2mm]
$W_H^+{}_{\mu }$  &  ${Z}_{\nu }$  &  $\Phi^-{}_{}$  &  $\Phi^p{}_{}$   & 
        $-\frac{ i e^2  \cdot\sqrt{2}}{ c_w s_w^2  f^2 }\varepsilon_{\mu \nu \rho \sigma} \big( (2-3 s_w ^2)p_2^\rho p_3^\sigma - (2-3 s_w ^2)p_1^\rho p_3^\sigma $ \\[2mm]
  &&&& $- s_w^2 p_1^\rho p_4^\sigma + s_w^2 p_2^\rho p_4^\sigma +2 (1-2 s_w ^2)p_3^\rho p_4^\sigma \big)$\\[2mm]
$W_H^-{}_{\mu }$  &  ${Z}_{\nu }$  &  $\Phi^+{}_{}$  &  $\Phi^0{}_{}$   & 
        $-\frac{ e^2  \cdot\sqrt{2}}{ c_w s_w^2  f^2 }\varepsilon_{\mu \nu \rho \sigma} \big( s_w^2 p_1^\rho p_4^\sigma - s_w^2 p_2^\rho p_4^\sigma - (2-3 s_w ^2)p_2^\rho p_3^\sigma $ \\[2mm]
  &&&& $+ (2-3 s_w ^2)p_1^\rho p_3^\sigma -2 (1-2 s_w ^2)p_3^\rho p_4^\sigma \big)$\\[2mm]
$W_H^-{}_{\mu }$  &  ${Z}_{\nu }$  &  $\Phi^+{}_{}$  &  $\Phi^p{}_{}$   & 
        $\frac{ i e^2  \cdot\sqrt{2}}{ c_w s_w^2  f^2 }\varepsilon_{\mu \nu \rho \sigma} \big( s_w^2 p_1^\rho p_4^\sigma - s_w^2 p_2^\rho p_4^\sigma - (2-3 s_w ^2)p_2^\rho p_3^\sigma $\\
    &&&& $+ (2-3 s_w ^2)p_1^\rho p_3^\sigma -2 (1-2 s_w ^2)p_3^\rho p_4^\sigma \big)$\\[2mm] 
%
${Z}_{\mu }$  &  $Z_H{}_{\nu }$  &  $\Phi^0{}_{}$  &  $\Phi^p{}_{}$   & 
        $2\frac{ i e^2 }{ c_w s_w^2  f^2 }\varepsilon_{\mu \nu \rho \sigma} \big(p_1^\rho p_3^\sigma -p_2^\rho p_3^\sigma -p_1^\rho p_4^\sigma +p_2^\rho p_4^\sigma +2p_3^\rho p_4^\sigma \big)$\\ \hline
\end{tabular}
\end{center}
\vspace{-1ex}
\mycaption{{\rm (Continuation of Tab.~\ref{tab:2gb2s})}
T-parity violating vertices with two scalars and two gauge bosons. 
The momenta $p_{1,2,3,4}$ correspond to the particle in the first, second, third and fourth column,
respectively.
{\rm Continued in Tab.~\ref{tab:2gb2s3}}}
\label{tab:2gb2s2}
\end{table}
\clearpage

\begin{table}[tp]
\begin{center}
\begin{tabular}{|cccc|c|} \hline
\multicolumn{4}{|l|}{\textbf{Particles}} & \textbf{Vertices} \\ \hline
$W_H^-{}_{\mu }$  &  ${Z}_{\nu }$  &  $\Phi^{++}{}_{}$  &  $\Phi^-{}_{}$   & 
        $-2\frac{ e^2 }{ c_w s_w^2  f^2 }\varepsilon_{\mu \nu \rho \sigma} \big( (2- s_w ^2)p_1^\rho p_4^\sigma - (2- s_w ^2)p_2^\rho p_4^\sigma $ \\[2mm]
  &&&& $+ s_w^2 p_2^\rho p_3^\sigma - s_w^2 p_1^\rho p_3^\sigma +2p_3^\rho p_4^\sigma \big)$\\[2mm]
${Z}_{\mu }$  &  $Z_H{}_{\nu }$  &  $\Phi^+{}_{}$  &  $\Phi^-{}_{}$   & 
        $6\frac{ c_w e^2 }{ s_w^2  f^2 }\varepsilon_{\mu \nu \rho \sigma} \big(p_1^\rho p_4^\sigma -p_2^\rho p_4^\sigma -p_1^\rho p_3^\sigma +p_2^\rho p_3^\sigma -2p_3^\rho p_4^\sigma \big)$\\[2mm]
${Z}_{\mu }$  &  $Z_H{}_{\nu }$  &  $\Phi^{++}{}_{}$  &  $\Phi^{--}{}_{}$   & 
        $2\frac{ (1-2 s_w ^2) e^2 }{ c_w s_w^2  f^2 }\varepsilon_{\mu \nu \rho \sigma} \big(p_1^\rho p_4^\sigma -p_2^\rho p_4^\sigma -p_1^\rho p_3^\sigma +p_2^\rho p_3^\sigma -2p_3^\rho p_4^\sigma \big)$\\[2mm]
$W^+_{\mu }$  &  $W_H^-{}_{\nu }$  &  $\Phi^+{}_{}$  &  $\Phi^-{}_{}$   & 
        $-2\frac{ e^2 }{ s_w^2  f^2 }\varepsilon_{\mu \nu \rho \sigma} \big(p_2^\rho p_4^\sigma -p_1^\rho p_4^\sigma +p_1^\rho p_3^\sigma -p_2^\rho p_3^\sigma +2p_3^\rho p_4^\sigma \big)$\\[2mm]
$W^+_{\mu }$  &  $W_H^-{}_{\nu }$  &  $\Phi^{++}{}_{}$  &  $\Phi^{--}{}_{}$   & 
        $2\frac{ e^2 }{ s_w^2  f^2 }\varepsilon_{\mu \nu \rho \sigma} \big(3p_1^\rho p_4^\sigma +p_2^\rho p_3^\sigma -p_1^\rho p_3^\sigma -3p_2^\rho p_4^\sigma -4p_3^\rho p_4^\sigma \big)$\\[2mm]
$W^+_{\mu }$  &  $W_H^-{}_{\nu }$  &  $\Phi^0{}_{}$  &  $\Phi^0{}_{}$   & 
        $-2\frac{ e^2 }{ s_w^2  f^2 }\varepsilon_{\mu \nu \rho \sigma} \big(p_1^\rho p_3^\sigma -p_2^\rho p_3^\sigma +p_1^\rho p_4^\sigma -p_2^\rho p_4^\sigma \big)$\\[2mm]
$W^+_{\mu }$  &  $W_H^-{}_{\nu }$  &  $\Phi^0{}_{}$  &  $\Phi^p{}_{}$   & 
        $-4\frac{ i e^2 }{ s_w^2  f^2 }\varepsilon_{\mu \nu \rho \sigma} \big(p_2^\rho p_3^\sigma -p_2^\rho p_4^\sigma +p_1^\rho p_4^\sigma -p_1^\rho p_3^\sigma -2p_3^\rho p_4^\sigma \big)$\\[2mm] \hline 
\end{tabular}
\end{center}
\vspace{-1ex}
\mycaption{{\rm (Continuation of Tab.~\ref{tab:2gb2s2})} T-parity violating vertices vertices  two scalars and two gauge bosons. 
The momenta $p_{1,2,3,4}$ correspond to the particle in the first, second, third and fourth column,
respectively.
}
\label{tab:2gb2s3}
\end{table}

\end{document}